\documentclass[11pt]{article}
\usepackage{eurosym}
\usepackage{graphicx}
\usepackage[utf8]{inputenc}
\usepackage[T1]{fontenc}
\usepackage{indentfirst}
\usepackage[margin=0.6in,nomarginpar]{geometry}
\usepackage[final]{hyperref}
\usepackage{amsmath}
\usepackage{hyperref}
\usepackage{cite}
\usepackage{subcaption}
\usepackage{caption}
\usepackage{amssymb}
\usepackage{multirow}
\usepackage[table]{xcolor}
\usepackage{orcidlink}
\usepackage{multirow,stackengine}
\usepackage{float}
\hypersetup{colorlinks=true, linkcolor=blue, citecolor=blue, urlcolor=blue}
\begin{document}

\title{Nonlinear Magnetically Charged Black Holes with Phantom Global Monopoles: Thermodynamics, Geodesics, Quasinormal Modes, and Grey-Body Factors}
\author{B. Hamil \orcidlink{0000-0002-7043-6104} \thanks{%
hamilbilel@gmail.com } \\
Laboratoire de Physique Math\'{e}matique et Subatomique,\\
Facult\'{e} des Sciences Exactes, Universit\'{e} Constantine 1 Frères Mentouri, Constantine,
Algeria. \and B. C. L\"{u}tf\"{u}o\u{g}lu 
\orcidlink{Orcid ID :
0000-0001-6467-5005} \thanks{%
bekir.lutfuoglu@uhk.cz (Corresponding author) } \\
Department of Physics, Faculty of Science, University of Hradec Kralove, \\
Rokitanskeho 62/26, Hradec Kralove, 500 03, Czech Republic. }
\date{\today }
\maketitle

\begin{abstract}
We study the properties of a nonlinear magnetic-charged black hole in the presence of a phantom global monopole. By incorporating nonlinear electrodynamics (NLE) and exotic scalar fields, we derive an exact black hole solution and analyze its geometric structure, causal properties, and thermodynamic behavior. We examine how the presence of a phantom global monopole modifies the black hole’s Hawking temperature, entropy, and stability conditions, revealing significant changes in its phase structure. Additionally, we investigate the geodesic motion of test particles. The quasinormal mode (QNM) spectrum is computed using the WKB approximation and Pöschl-Teller potential method, providing insights into the perturbative stability of the system. Furthermore, we analyze the grey-body factors that characterize radiation emission, highlighting their dependence on black hole parameters. Our findings indicate that the interplay between phantom energy, NLE, and global monopoles introduces observable deviations in strong-field astrophysical phenomena. These results offer potential signatures for testing modified gravity theories and contribute to a deeper understanding of black hole physics in exotic field environments.  
\end{abstract}


\section{Introduction}

Black holes in modified gravity theories and exotic matter environments exhibit profound deviations from general relativity (GR), offering unique opportunities to probe high-energy astrophysics and quantum gravity effects \cite{Moffat2015, Mureika2016, Cardoso2019, Mann2022, Sheikamadi2023}. While classical black hole solutions in GR are well established, alternative theories predict significant modifications in strong-field regimes. These deviations can manifest in astrophysical observations such as black hole shadows, gravitational waves, and high-energy emissions, providing crucial insights into physics beyond GR. Additionally, the presence of exotic fields and nonlinear electrodynamics (NLE) around black holes alters their thermodynamic stability, causal structure, and radiation properties. Investigating these effects is essential for distinguishing modified gravity scenarios from standard astrophysical processes. For a comprehensive review of these modifications and their implications, see \cite{Cardoso2019} and references therein.

Among the exotic fields that may alter black hole physics, phantom energy is of particular interest due to its ability to violate the null energy condition, leading to counterintuitive consequences \cite{Caldwell2002}. Unlike conventional dark energy models, phantom fields have an equation of state parameter less than $-1$, which can significantly impact black hole thermodynamics and stability. Observational data have not ruled out phantom energy \cite{Melchiorri2003}, making it an active research topic across various branches of physics. Notably, phantom energy accretion reduces black hole mass, potentially violating cosmic censorship and the generalized second law of thermodynamics \cite{Babichev2004, Izquierdo2006, Sadjadi2007}. In Reissner-Nordström (RN)-like black holes, this process increases the charge-to-mass ratio beyond unity, modifying standard black hole dynamics \cite{Jamil2011b, Jardim2012}. Furthermore, phantom fields affect gravitational lensing properties \cite{Eiroa2013}, strong lensing effects, and light propagation \cite{AzregAinou2013, Gyulchev2013, Ding2013}. Beyond gravitational effects, phantom fields alter black hole thermodynamics by increasing temperature as mass decreases, influencing the evaporation process via Hawking radiation \cite{Jamil2012}. However, the stability of phantom-dominated black holes remains uncertain, as negative-energy fluxes can introduce instabilities in the event horizon structure \cite{Jamil2011a, Panah2024, Eslam2023, Jafarzade2024,  Hamil2025b}. Further studies in various black hole models, including BTZ black holes in commutative \cite{Panah2024} and noncommutative \cite{Hamil2025b} spaces, have highlighted the need for a deeper investigation into their thermodynamic fate and role in an accelerating universe.

Another fundamental extension of GR involves NLE, which modifies Maxwell’s theory to account for strong-field effects \cite{AyonBeato1998, AyonBeato1999}. Such modifications arise naturally in high-energy regimes, particularly around magnetized black holes or intense electromagnetic fields \cite{Cataldo2000, Bronnikov2001}. Several studies indicate that NLE can lead to regular black hole solutions, mitigating singularities while preserving essential thermodynamic properties \cite{Balart2014, Kruglov2017, Hendi2014}. The thermodynamics of black holes within NLE frameworks have been extensively analyzed, revealing rich phase structures, critical behavior, and modified heat capacities \cite{Gonzalez2009, Kuang2018, Paul2024, Awal2024}. However, when combined with exotic fields such as phantom energy, NLE introduces additional complexities affecting black hole stability, causal structure, and accretion dynamics \cite{Kruglov2016, Sudhanshu2024}. Recent research has explored magnetically charged black holes in NLE coupled to phantom fields, uncovering novel regular solutions and black bounce geometries \cite{Canate2022}. Additionally, investigations into rotating and nonlinearly charged black holes with anisotropic matter fields have demonstrated significant deviations from standard GR solutions \cite{Li2025}. These modifications influence observable phenomena, including black hole shadows, gravitational lensing, and quasinormal modes (QNMs), providing potential avenues for testing deviations from classical GR \cite{ Kruglov2023}. Lower-dimensional studies, such as those on BTZ black holes, further highlight the unique thermodynamic and geometric properties introduced by NLE \cite{Paul2024}. As research continues to explore the interplay between NLE and unconventional matter fields, these studies offer valuable insights into quantum gravity corrections, alternative compact objects, and broader modified gravity theories.

An intriguing and less-explored aspect of modified gravity is the role of topological defects, particularly global monopoles, in black hole physics. Global monopoles arise from the spontaneous symmetry breaking of a global \(O(3)\) symmetry to \(O(2)\) and differ from gauge monopoles due to their long-range gravitational effects \cite{Barriola1989}. Unlike gauge monopoles, which have localized energy densities, global monopoles introduce a conical deficit angle that modifies spacetime on large scales. Their gravitational influence has been widely studied, including effects on vacuum polarization \cite{Yu1994}, quantum singularities \cite{Chen2013}, and general relativistic solutions \cite{Sharif2015, Beato2021}. When coupled with black holes, global monopoles significantly alter event horizon properties, geodesic motion, and thermodynamics, leading to distinct phenomena such as modified Hawking radiation \cite{Jiang2006, Jiang2006b, Chen2006, Gangopadhyay2008, Pu2017}, gravitational lensing effects \cite{Sharif2015, Zhang2014}, and deviations in accretion processes \cite{Ahmed2016, Xiang2011}. The presence of a global monopole also affects black hole shadows and other observational features in various modified gravity models \cite{Haroon2020, Anacleto2023}. Furthermore, the thermodynamic behavior of such black holes has been extensively examined, revealing critical phase transitions \cite{Soroushfar2020, Deng2018}, modified entropy relations \cite{Jing2013, Majeed2017}, and possible Planck-scale remnants \cite{Li2017}. Studies on charged and rotating black holes with global monopoles suggest substantial deviations from classical solutions, influencing stability and causal structure \cite{Jiang2006b, Gangopadhyay2008, Soroushfar2020}. Additional research on global monopoles in noncommutative geometries has shown modifications in quantum tunneling radiation and black hole evaporation \cite{Chen2006, Jiang2006b, Anacleto2023}. These findings highlight the complex interplay between global monopoles, exotic matter fields, and nonlinear electrodynamics, offering potential avenues to test deviations from GR through astrophysical observations.

Despite extensive research on phantom energy, nonlinear electrodynamics, and global monopoles individually, their combined impact on black hole physics remains largely unexplored. { This study aims to fill this gap by systematically investigating how their interplay affects black hole thermodynamics, stability, and observational characteristics. Specifically, we construct and analyze a regular black hole solution sourced by a nonlinear magnetic field and pierced by a phantom global monopole. This setup reveals rich physical behavior, including nontrivial modifications to the horizon structure, thermodynamic phase transitions, and the spectrum of scalar quasinormal modes. Our results demonstrate that the combined effects of phantom topology and nonlinear electrodynamics introduce distinctive signatures that could, in principle, be probed observationally. To our knowledge, this is the first detailed investigation of such a composite system, highlighting its theoretical novelty and potential phenomenological relevance}. This paper is structured as follows: Section~\ref{sec2} presents the derivation of the black hole solution and its geometric properties, while Section~\ref{sec3} analyzes its thermodynamic behavior. Section \ref{sec4} investigates the timelike geodesics and the motion of test particles in the black hole spacetime. Section~\ref{sec5} examines QNMs using the sixth-order WKB approximation, cross-checked with the Pöschl-Teller potential. Finally, Section~\ref{sec6} analyzes grey-body factors and their influence on radiation emission, before we conclude in Section~\ref{sec7} with a summary and future outlook.

\section{Spacetime Geometry and Black Hole Metric} \label{sec2}

We begin by investigating a static, spherically symmetric black hole embedded within a phantom global monopole. This configuration arises due to the spontaneous symmetry breaking of a triplet of phantom scalar fields associated with a global $O(3)$ symmetry. The presence of the global monopole modifies the action, which takes the form \cite{Barriola1989}:
\begin{equation} 
S=\frac{1}{16\pi }\int \sqrt{-g}Rd^{4}x+\int d^{4}x\sqrt{-g}\mathcal{L}^{M}, 
\end{equation}
where the matter Lagrangian is given by
\begin{equation} \mathcal{L}^{M}=\frac{\xi }{2}\partial _{\mu }\chi ^{a}\partial ^{\mu }\chi ^{a}-\frac{\lambda }{4}\left( \chi ^{a}\chi ^{a}-\eta ^{2}\right) ^{2}. \qquad a=1,2,3. \end{equation}
Here, $\chi ^{a}=\eta h\left( r\right) \frac{x^{a}}{r}$ represents a scalar field triplet, where $h\left( r\right) $ is a dimensionless function, $\eta $ is the energy scale of symmetry breaking, and $\lambda$ is a coupling constant. When $\xi=1$, the setup corresponds to a conventional global monopole, characterized by a scalar field with positive kinetic energy. However, in the case where $\xi=-1$, the kinetic energy becomes negative, leading to the formation of a phantom global monopole.

To analyze a static, spherically symmetric black hole in the presence of a phantom global monopole, we consider the following line element:
\begin{equation}
ds^{2}=-B\left( r\right) dt^{2}+A\left( r\right) dr^{2}+r^{2}\left( d\theta^{2}+\sin ^{2}\theta d\phi ^{2}\right),  \label{x3}
\end{equation}
which describes the gravitational field around the black hole. In this metric background, the field equations for the triplet scalar field $\chi ^{a}$ reduce to a single differential equation governing $h(r)$ \cite{Chen2013}:
\begin{equation}
\xi \frac{h^{\prime \prime }}{A}+\left[ \frac{1}{Ar}+\frac{1}{2B}\left( 
\frac{B}{A}\right) \right] \xi h^{\prime }-\frac{2\xi h}{r^{2}}-\lambda \eta
^{2}h\left( h^{2}-1\right) =0.
\end{equation}
{ Following the approach of Barriola and Vilenkin  \cite{Barriola1989},  we assume that the scalar field function $h(r)$ quickly approaches unity outside the monopole core, since $h(r)$ grows linearly for $r< \delta$ and decays exponentially toward unity once $r> \delta$, where $\delta \sim (\eta \sqrt{\lambda})^{-1}$ defines the core radius of the global monopole.} Under this assumption, the energy-momentum tensor associated with the global monopole takes the form:

\begin{equation}
T_{t}^{t\left( \mathrm{M}\right) }=T_{r}^{r\left( \mathrm{M}\right) }=\frac{%
\xi \eta ^{2}}{r^{2}}, \quad T_{\theta }^{\theta \left( \mathrm{M}%
\right) }=T_{\phi }^{\phi \left( \mathrm{M}\right) }=0.
\end{equation}

On the other hand, when Einstein gravity is coupled to an NLE field in the presence of a phantom global monopole,  the total action can be written as:
\begin{equation} 
\mathcal{S}=\frac{1}{16\pi }\int \sqrt{-g}Rd^{4}x+\int d^{4}x\sqrt{-g}\Big(\mathcal{L}^{M}-\frac{1}{4\pi} \mathcal{L}^{charge}\Big),\label{6} 
\end{equation}
where the NLE Lagrangian density is given by:
\begin{equation}
\mathcal{L}^{charge}=\frac{3m}{\left\vert Q\right\vert ^{3}}\frac{\left(
2Q^{2}F\right) ^{3/2}}{\left[ 1+\left( 2Q^{2}F\right) ^{4/3}\right] ^{2}}.\label{8}
\end{equation}
Here, $F=\frac{1}{4}F_{\mu \nu }F^{\mu \nu }$, while $m$ and $Q$ correspond to the mass and magnetic charge of the system, respectively. Here, we introduce the magnetic charge $Q$ in the Lagrangian explicitly, since this mechanism is solution-dependent. For the present NLE case, this strategy is adopted and it follows reconstruction of the Lagrangian from a solution. A more general approach would be to introduce such a parameter and then relate it later; however, our choice is for analytic tractability and to elucidate the role played by the phantom monopole. In this work, we adopt a NLE Lagrangian that supports magnetically charged black hole solutions with desirable properties. The form given in Eq. (\ref{8}) is inspired by regular black hole models such as the well-known Hayward solution, which has been shown to arise from specific NLE frameworks. Our aim  is to investigate how this well-motivated NLE framework interacts with phantom global monopole fields, rather than to explore the most general possible NLE formulation.
We assume that the scalar and NLE fields are minimally coupled, i.e., they interact only through the gravitational sector, and no direct coupling term is introduced between them. This choice is deliberate and well-motivated. The scalar field generates a global conical deformation in the asymptotic geometry, characteristic of global monopoles \cite{Barriola1989, Sharif2015}, while the NLE field dominates the near-horizon behavior and is responsible for softening the central singularity \cite{Hendi2014, Balart2014}. By keeping the two sectors minimally coupled, we ensure that their distinct effects remain analytically separable and theoretically transparent. Introducing a direct scalar-electromagnetic coupling would introduce new coupling parameters and complicate the interpretation of results, obscuring the independent roles of the scalar and electromagnetic sources.

Applying the variational principle from Eq. (\ref{6}), the field equations governing the system take the form:
\begin{equation}
G_{\mu }^{\nu }=2\left( \frac{\partial \mathcal{L}^{charge}}{\partial F}%
F_{\mu \alpha }F^{\alpha \nu }-\delta _{\mu }^{\nu }\mathcal{L}%
^{charge}\right) +8\pi T_{\mu }^{\nu \left( \mathrm{M}\right) },  \label{EE}
\end{equation}
\begin{equation}
\nabla _{\mu }\left( \frac{\partial \mathcal{L}^{charge}}{\partial F}F^{\nu\mu }\right) =0,
\end{equation}

Next, we adopt a spherically symmetric spacetime ansatz of the form:
\begin{equation}
ds^{2}=-f\left( r\right) dt^{2}+\frac{1}{f\left( r\right) }%
dr^{2}+r^{2}\left( d\theta ^{2}+\sin ^{2}\theta d\phi ^{2}\right),  
\end{equation}
where
\begin{equation}
    f\left( r\right) =1-\frac{2M\left( r\right) }{r}.
\end{equation}
To describe the electromagnetic field, we employ the Maxwell field ansatz \cite{Beato2021}:
\begin{equation}
F_{\mu \nu }=\left( \delta _{\mu }^{\theta }\delta _{\nu }^{\phi }-\delta
_{\nu }^{\theta }\delta _{\mu }^{\phi }\right) B\left( r,\theta \right)
=\left( \delta _{\mu }^{\theta }\delta _{\nu }^{\phi }-\delta _{\nu
}^{\theta }\delta _{\mu }^{\phi }\right) Q\sin \theta ,
\end{equation}
which directly leads to:
\begin{equation}
F=\frac{Q^{2}}{2r^{4}}.
\end{equation}
The electromagnetic invariant remains strictly positive throughout the spacetime. This ensures that despite the non-integer exponents appearing in Eq. (\ref{8}), the Lagrangian maintains both reality and regularity everywhere outside the origin. We note, however, that this form of the Lagrangian may become ill-defined in more general scenarios where $F$ could change sign, and this is a limitation of the specific ansatz we consider. Using the previously derived equations, the time component of Eq. (\ref{EE}) simplifies to:
\begin{equation}
-\frac{2}{r^{2}}\frac{dM\left( r\right) }{dr}=-\frac{6mQ^{3}}{\left(
r^{3}+Q^{3}\right) ^{2}}+8\pi \frac{\xi \eta ^{2}}{r^{2}}.
\end{equation}
Integrating this expression, we obtain the metric function that describes a black hole with a NLE field in the presence of a phantom global monopole:
\begin{equation}
f(r)=1-8\pi \xi \eta ^{2}-\frac{2mr^{2}}{r^{3}+Q^{3}}.\label{114}
\end{equation}
We note that in the absence of the scalar field,  the solution in Eq. (\ref{114}) reduces precisely to the Hayward regular black hole \cite{65}. Later the Hayward black hole was described by NLE theory in \cite{66}. In the special case where $\xi \rightarrow 0$, this solution reduces to a black hole with a nonlinear magnetic charge. On the other hand, setting $Q\rightarrow 0$ retrieves the Schwarzschild black hole solution with a phantom global monopole \cite{Sharif2015}. We note that our choice of a regular NLE Lagrangian guarantees well-behaved field invariants and stress–energy if one extends the model toward small radii or matches to an interior solution. However, for the exterior analysis ($r \geq r_c$) presented here, the qualitative results remain unchanged for other standard NLE models (e.g., Maxwell, Born–Infeld, logarithmic), with only minor quantitative differences. This confirms that our main conclusions are robust against the specific NLE choice. It is crucial to highlight that NLE models, such as the square-root model—have been shown to induce angular deficits or excesses without invoking scalar fields \cite{67}; however, these effects arise purely from the electromagnetic sector. In contrast, our work investigates the combined influence of two physically distinct sources: a phantom global monopole, which acts as a topological defect and modifies the large-scale structure of spacetime, and a nonlinear magnetic field, which governs the near-horizon geometry. These fields are minimally coupled, interacting only through gravity, which allows us to isolate their individual contributions while studying their collective impact on black hole physics. Although the inclusion of a phantom monopole renders the spacetime non–asymptotically flat, in this work we restrict attention to the exterior region $r \ge r_c$ relevant to the global–monopole setup. This setup provides a novel framework for exploring the interplay between topological charge and regularization mechanisms, with potential implications for black hole thermodynamics, quasinormal mode spectra, and gravitational wave phenomenology.

To investigate the curvature behavior of this spacetime, we compute the Ricci scalar $\mathbf{R}$ and the Kretschmann scalar $\mathbf{K}$, given by:
\begin{equation}
\mathbf{R}=\frac{16\pi \eta ^{2}\xi }{r^{2}}+\frac{4m\left(
6Q^{6}r^{2}-3Q^{3}r^{5}\right) }{\left( Q^{3}+r^{3}\right) ^{3}r^{2}},
\end{equation}%
\begin{equation}
\mathbf{K}=\frac{48m^{2}\left(
2Q^{12}-2Q^{9}r^{3}+18Q^{6}r^{6}-4Q^{3}r^{9}+r^{12}\right) }{\left(
Q^{3}+r^{3}\right) ^{6}}+\frac{128\pi \eta ^{2}m\xi }{r^{2}\left(
Q^{3}+r^{3}\right) }+\frac{256\pi ^{2}\eta ^{4}\xi ^{2}}{r^{4}}. \label{Kretschmann}
\end{equation}
As Eq.~\eqref{Kretschmann} shows, the Kretschmann scalar diverges as $r \to 0$, which means that the spacetime is not globally regular at the origin. In our analysis we only consider the exterior region $r \ge r_c$, where the geometry remains regular. 

In Table \ref{x} we include three NLE models that generate regular black holes, namely the Hayward, Bardeen, and Ayon-Beato-Garcia (ABG) spacetimes. Their metric functions, together with the corresponding curvature properties, are summarized there. We find that for $\eta =0$ all these models are regular at the origin, whereas for $\eta \neq 0$ the pointlike monopole contribution dominates and inevitably leads to a
curvature divergence.
\begin{table}[H]
\renewcommand{\arraystretch}{1.5}
\centering%
\begin{tabular}{|l|l|l|l|}
\hline\hline
\rowcolor{lightgray} NED model & Metric function & $\lim\limits_{r\rightarrow 0}$ $\mathbf{K}$ ($\eta =0$%
) & \multicolumn{1}{|l|}{$\lim\limits_{r\rightarrow 0}$ $\mathbf{K}$ ($\eta \neq 0$)
} \\ \hline
Hayward & $1-8\pi \xi \eta ^{2}-\frac{2mr^{2}}{r^{3}+Q^{3}}$ & $\frac{96m^{2}%
}{Q^{6}}$ & $\infty $ \\ 
Bardeen & $1-8\pi \xi \eta ^{2}-\frac{2mr^{2}}{\left( r^{2}+Q^{2}\right)
^{3/2}}$ & $\frac{96m^{2}}{Q^{2}}$ & $\infty $ \\ 
ABG & $1-8\pi \xi \eta ^{2}-\frac{2mr^{2}}{\left( r^{2}+Q^{2}\right) ^{3/2}}+%
\frac{Q^{2}r^{2}}{\left( r^{2}+Q^{2}\right) ^{2}}$ & $\frac{24\left(
2m-Q\right) ^{2}}{Q^{6}}$ & $\infty$\\
\hline
\end{tabular}
\caption{Comparison of Hayward, Bardeen, and ABG BHs: metric functions and Kretschmann scalar limits with and without monopole.}
\label{x}
\end{table}

To further understand the influence of the phantom global monopole on the black hole structure, we plot the metric function $f(r)$ for different values of the symmetry-breaking scale $\eta$. This allows us to examine how the presence of the monopole affects the event horizon and overall spacetime geometry.

\begin{figure}[H]
\begin{minipage}[t]{0.33\textwidth}
        \centering
            \includegraphics[width=\textwidth]{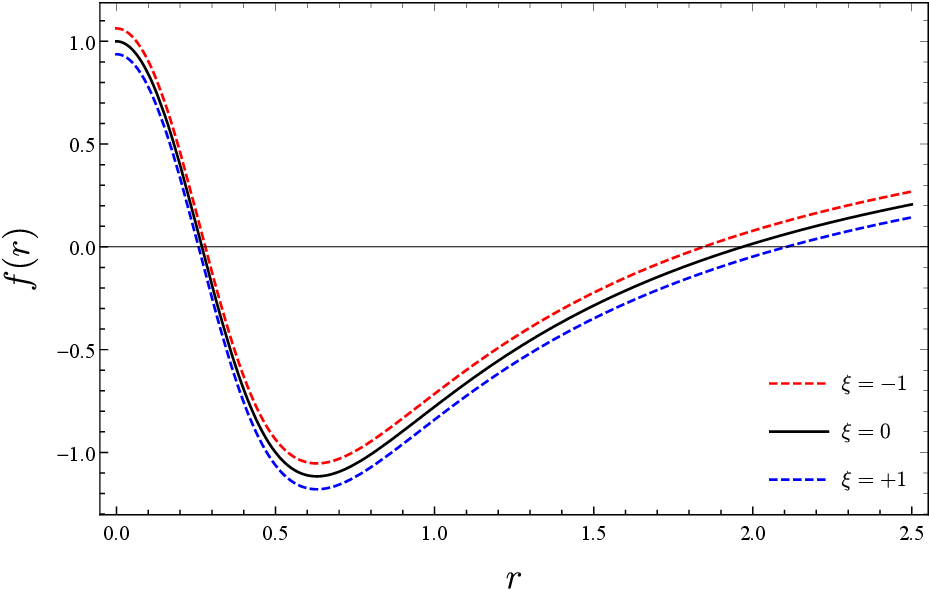}
            \subcaption{$\eta=0.05$}
         \label{fig:me1}
\end{minipage}%
\begin{minipage}[t]{0.33\textwidth}
        \centering
               \includegraphics[width=\textwidth]{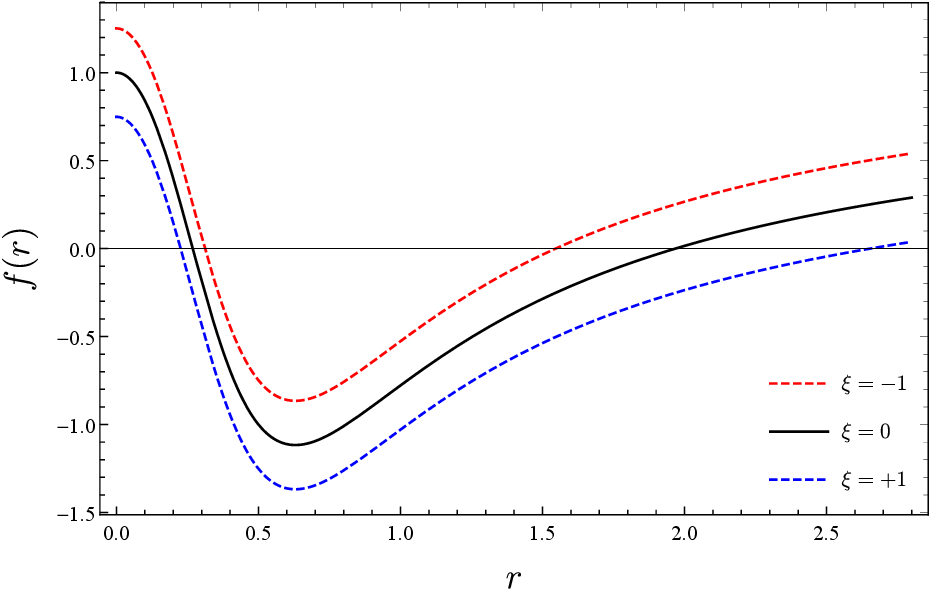}
                \subcaption{$\eta=0.1$}
       \label{fig:me2}
   \end{minipage}%
\begin{minipage}[t]{0.33\textwidth}
        \centering
                \includegraphics[width=\textwidth]{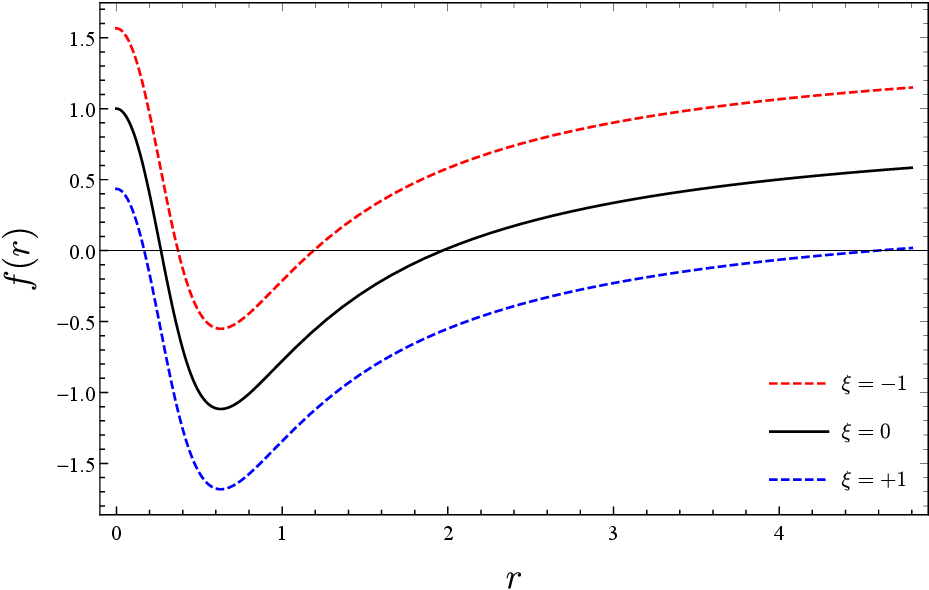}
                \subcaption{$\eta=0.15$}
         \label{fig:me3}
   \end{minipage}
\caption{Behavior of the metric function $f(r)$ for different values of the monopole parameter $\eta$, with fixed charge $Q =0.5$ and mass $m =1$.}
\label{fig:metri}
\end{figure}

\noindent As observed in Figure \ref{fig:metri}, increasing the symmetry-breaking parameter $\eta$ leads to a more pronounced deviation of the metric function $f(r)$. Specifically, higher values of $\eta$ result in a deeper gravitational potential, shifting the position of the event horizon outward. This is a direct consequence of the monopole's additional energy contribution, which effectively modifies the gravitational field. 

Moreover, the presence of the parameter $\xi$ (implicitly affecting $\eta$) influences the global behavior of the solution, indicating that a sufficiently strong monopole field can significantly alter black hole thermodynamics and causal structure. In particular, the limiting behavior at large $r$ remains consistent, although the spacetime is not asymptotically flat due to the solid-angle (conical) deficit induced by the monopole. However, near the black hole horizon, the modifications become more pronounced, potentially impacting observational signatures such as gravitational lensing and black hole shadows.

In this work, we adopt a specific form of NLE that reproduces the regular black hole solution proposed by Hayward \cite{65} and later derived from a magnetic NLE source \cite{66}. This choice ensures central regularity and allows for analytical control. We set the Schwarzschild mass parameter to zero, following the original Hayward formulation, where the effective mass emerges dynamically from the nonlinear magnetic sector. This assumption is particularly useful in our setup, as it allows us to isolate and analyze the gravitational interplay between the phantom global monopole and the NLE sector. While the phantom field alters the asymptotic structure through a topological angular deficit, the NLE Lagrangian governs the near-horizon geometry and ensures regularity at the core. The resulting spacetime, although not globally regular or asymptotically flat, offers a controlled framework to explore the combined geometric and thermodynamic effects of these two physically distinct sources.
\section{Thermodynamics} \label{sec3}
In this section, we focus on exploring the thermodynamic properties of the black hole by analyzing its Hawking temperature. The Hawking temperature is defined as
\begin{equation}
T=\frac{\kappa }{2\pi }=\frac{f^{\prime }\left( r_{H}\right) }{4\pi },
\label{deft}
\end{equation}%
where $\kappa$ represents the surface gravity at the event horizon. Although a global monopole introduces a solid-angle deficit, the Hawking temperature defined via the surface gravity, $T=\kappa/(2\pi)$ with $\kappa=\tfrac{1}{2}f'(r_H)$, is a local quantity at the horizon and remains valid. The angular deficit affects the horizon area (and thus the entropy), not the surface gravity.

To evaluate this temperature, we first determine the black hole mass $m$ in terms of the horizon radius $r_H$ by solving the condition $f(r_H) = 0$. This leads to the following expression:
\begin{equation}
m=\frac{r_{H}}{2}\left( 1-8\pi \xi \eta ^{2}\right) \left( 1+\frac{Q^{3}}{%
r_{H}^{3}}\right). \label{mass}
\end{equation}%
Substituting Eq. (\ref{mass}) into Eq. (\ref{deft}), we obtain the Hawking temperature:
\begin{equation}
T=\frac{\left( 1-8\pi \xi \eta ^{2} \right)}{4\pi r_{H}} \frac{\left( r_{H}^{3}-2Q^{3}\right) }{ \left( r_{H}^{3}+ Q^{3}\right) }.  \label{tem}
\end{equation}%
For the Hawking temperature to remain both real and positive, two conditions must be satisfied. The first constraint stems from the NLE field, establishing a lower bound on the event horizon radius $r_H$ in terms of the magnetic charge $Q$:
\begin{equation}
r_{H}\geq 2^{1/3}Q.
\end{equation}
The second constraint ensures the positivity of temperature when $\xi =1$, leading to an upper bound on the symmetry-breaking energy scale:
\begin{equation}
\eta <\frac{1}{\sqrt{8\pi }}.
\end{equation}
These constraints highlight the interplay between the black hole horizon, charge, and the presence of the phantom global monopole, significantly influencing the thermodynamic behavior of the system.

\begin{figure}[H]
\begin{minipage}[t]{0.33\textwidth}
        \centering
            \includegraphics[width=\textwidth]{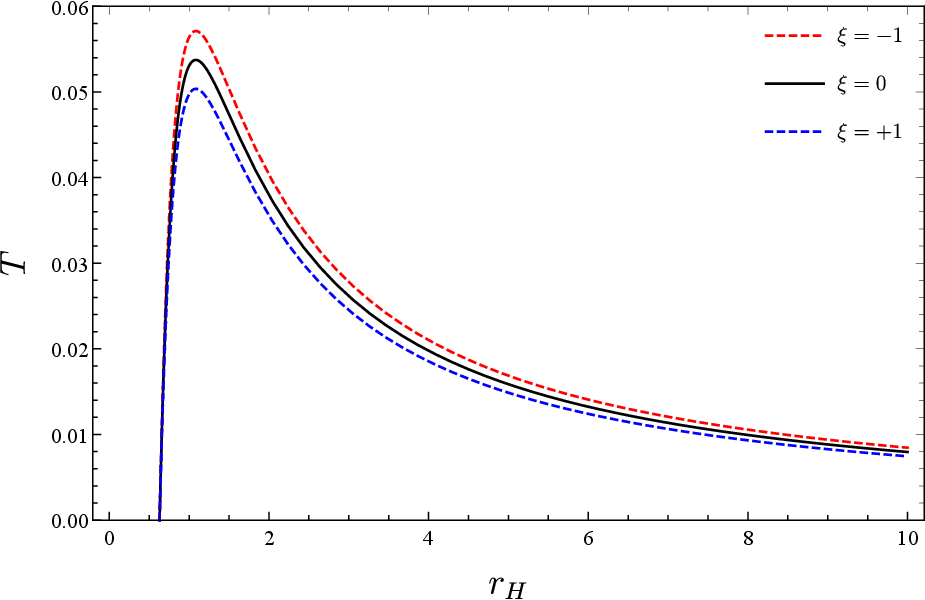}
            \subcaption{$\eta=0.05$}
         \label{fig:temp1}
\end{minipage}%
\begin{minipage}[t]{0.33\textwidth}
        \centering
               \includegraphics[width=\textwidth]{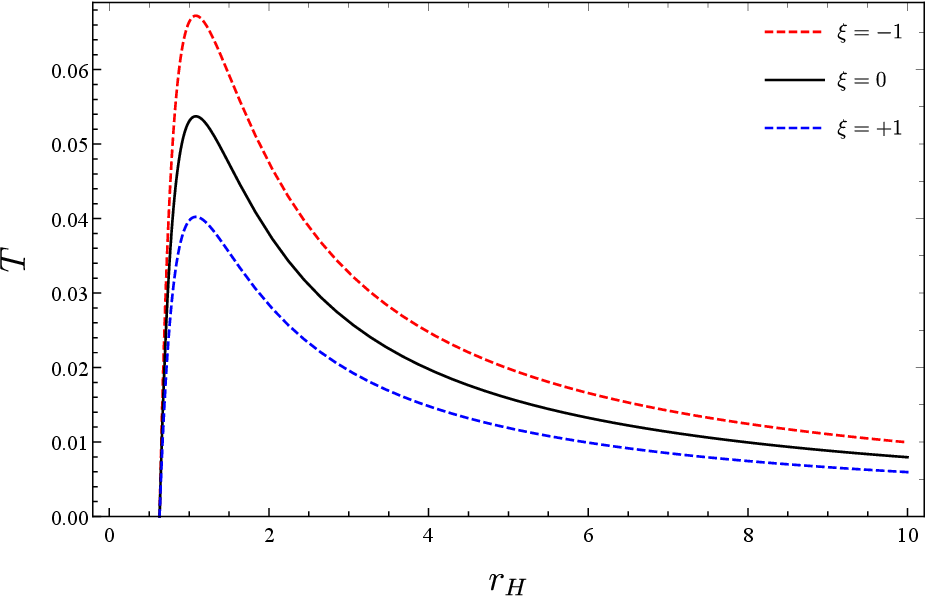}
                \subcaption{$\eta=0.1$}
       \label{fig:temp2}
   \end{minipage}%
\begin{minipage}[t]{0.33\textwidth}
        \centering
                \includegraphics[width=\textwidth]{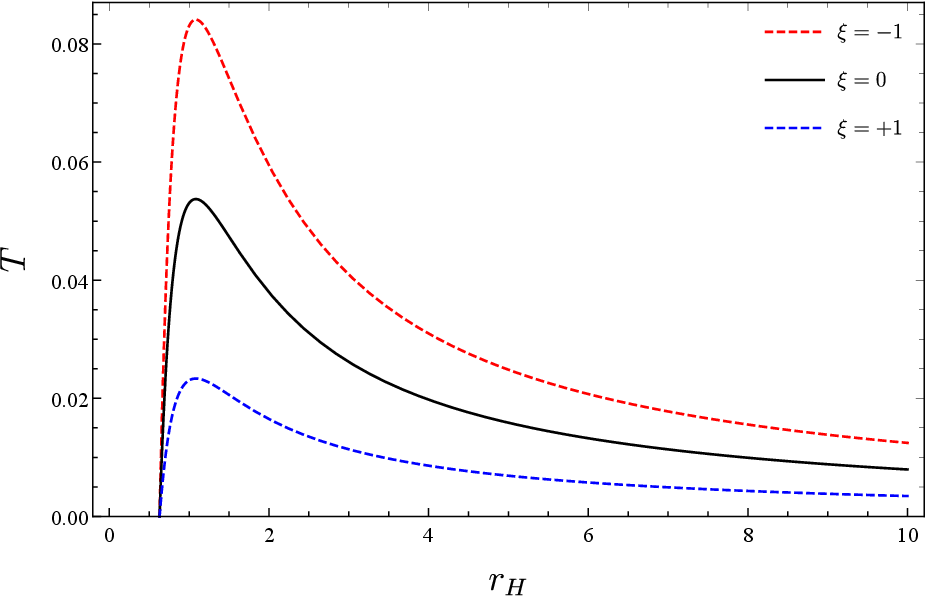}
                \subcaption{$\eta=0.15$}
         \label{fig:temp3}
   \end{minipage}
\caption{Hawking temperature $T$ as a function of the horizon radius $r_H$ for different values of the symmetry-breaking scale $\eta$. The black hole parameters are set to $Q =0.5$ and $m =1$. The effect of the phantom global monopole ($\xi=-1$) leads to an increase in temperature, whereas the ordinary global monopole ($\xi=+1$) lowers it.}
\label{fig:temperature}
\end{figure}

A significant observation is that the Hawking temperature vanishes at $r_{H}=2^{1/3}Q$, indicating that the black hole ceases to emit Hawking radiation. In Figure \ref{fig:temperature}, we depict the Hawking temperature of the nonlinear magnetic-charged black hole as a function of the horizon radius $r_{H}$ for three different cases: $\xi=-1,0,+1$. 

\noindent From this plot, we observe that the Hawking temperature decreases as the horizon radius increases, meaning that larger black holes tend to have lower temperatures. When $\xi =-1$, the Hawking temperature is higher compared to the other cases for a given $r_H$, implying that the presence of the phantom global monopole amplifies the black hole's temperature. Conversely, for $\xi =+1$, the temperature is the lowest among the three cases, confirming that while the phantom global monopole increases the black hole's temperature, the ordinary global monopole reduces it. Furthermore, the figure illustrates that during the black hole's evaporation process, the Hawking temperature initially rises, reaching a peak value $T_{\max }$ at a critical horizon radius $r_{H}^{\mathrm{c}}$. Beyond this point, the temperature rapidly declines, ultimately reaching zero. At this stage, the black hole stops radiating, and the pair-production processes occurring near $r_{H}^{\mathrm{c}}$ come to an end. Additionally, we note that as the energy scale of symmetry breaking $\eta$ increases, the Hawking temperature exhibits different behaviors in the two cases: it rises in the presence of the phantom global monopole but decreases when an ordinary global monopole is present.

We now turn our attention to the second thermodynamic quantity, entropy, which is defined by the relation:
\begin{equation}
dS=\frac{dm}{T}.
\end{equation}
Using the expression for the black hole mass (\ref{mass}) and the Hawking temperature (\ref{tem}), we obtain the following expression for the entropy:
\begin{equation}
S=\frac{A}{4}-\frac{4\pi ^{3/2}Q^{3}}{\sqrt{A}},  \label{ent}
\end{equation}
where $A=4\pi {(1 - 8\pi \xi \eta^2)} r_{H}^{2}$ is the physical horizon area rescaled by the solid–angle deficit due to a global monopole. It is noteworthy that the entropy does not exactly coincide with the standard Bekenstein--Hawking law. In particular, Eq.~ \eqref{ent} contains an additional correction term arising from NLE. The standard Bekenstein--Hawking form is recovered either when $\xi\!\to\!0$ and $Q\!\to\!0$ (the conventional case without a monopole and without NLE), or, approximately, in the large–area limit where the NLE correction becomes negligible.

In addition to the temperature and entropy analysis, we derive a first law and a Smarr-like formula for the nonlinearly magnetically charged black hole in the presence of a phantom global monopole. From Eq. (\ref{mass}), we express the black hole mass $m$ in terms of the horizon radius $r_{H}$, magnetic charge $Q$ and symmetry-breaking scale $\eta$. Differentiating $m$, we find
\begin{equation}
 dm=TdS+\Psi dQ+\Phi d\eta,
\end{equation}
where 
\begin{equation}
\Psi=\left. \frac{\partial m}{\partial Q}\right\vert _{r_{H},\eta }= \frac{3r_{H}}{2}\left( 1-8\pi \xi \eta
^{2}\right) \frac{Q^{2}}{r_{H}^{3}},   
\end{equation}
and
\begin{equation}
\Phi=\left. \frac{\partial m}{\partial {\eta}}\right\vert _{r_{H},Q }=-r_{H}\left( 1+\frac{Q^{3}}{r_{H}^{3}}%
\right) 8\pi \xi \eta,    
\end{equation}
represent the generalized potentials conjugate to $Q$ and $\eta$, respectively. These terms account for the contributions of the NLE and the phantom global monopole to the black hole’s energy content. Furthermore, using Euler’s theorem for homogeneous functions, we obtain a Smarr-type formula:
\begin{equation}
m=2TS+\Psi Q-\Phi \eta.    
\end{equation}

Next, we investigate the thermodynamic stability of the black hole by analyzing its heat capacity. The heat capacity is defined as:

\begin{equation}
C=T\left( \frac{\partial S}{\partial T}\right),  \label{Heat}
\end{equation}
and can be expressed, using Eqs. (\ref{tem}), (\ref{ent}), and (\ref{Heat}), in the following form:
\begin{equation}
C=-\frac{2\pi \left( r_{H}^{3}-2Q^{3}\right) \left( r_{H}^{3}+Q^{3}\right)
^{2}}{r_{H}^{7}-10Q^{3}r_{H}^{4}-2Q^{6}r_{H}}.
\end{equation}

\begin{figure}[H]
    \centering
    \includegraphics[width=0.5\linewidth]{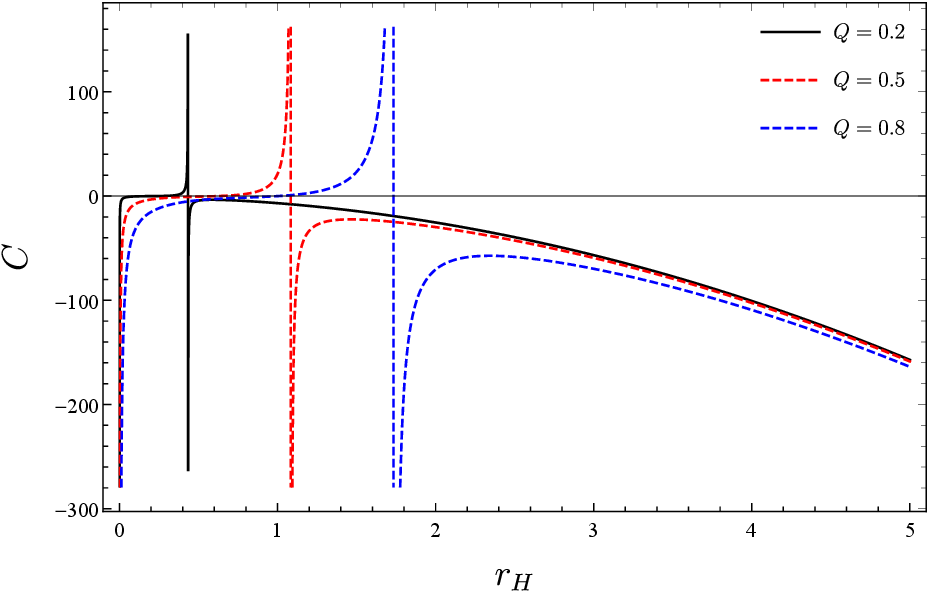}
    \caption{Heat capacity $C$ as a function of the horizon radius $r_H$ for the nonlinear magnetic-charged black hole. The discontinuity in $C$ indicates the presence of a second-order phase transition, separating an unstable region ($C<0$) from a stable region ($C>0$).}
    \label{fig:heat}
\end{figure}

\noindent In Figure \ref{fig:heat}, we plot the heat capacity of the nonlinear magnetic-charged black hole as a function of the horizon radius $r_{H}$. 

\noindent From this plot, we observe the presence of a discontinuity for each value of the magnetic charge parameter $Q$. Physically, this discontinuity signifies that the black hole undergoes a second-order phase transition. This transition separates an unstable phase, where $C<0$, from a stable phase, where $C>0$. Specifically, for smaller black holes (low $r_H$), the negative heat capacity indicates that they lose mass and evaporate due to Hawking radiation. However, beyond the critical point where $C$ changes sign, the black hole enters a stable phase, implying that it can remain in equilibrium with its surroundings.

To gain deeper insight into the thermal phase transition, we now analyze the Gibbs free energy $G$, which is defined as:
\begin{equation}
G=m-TS.
\end{equation}
Substituting the expressions for the black hole mass (\ref{mass}), Hawking temperature (\ref{tem}), and entropy (\ref{ent}), we obtain the Gibbs free energy in the following form:
\begin{equation}
G=\frac{\left( 1-8\pi \eta ^{2}\xi \right) \left(
r_{H}^{6}+8Q^{3}r_{H}^{3}-2Q^{6}\right) }{4\,r_{H}^{2}\left(
r_{H}^{3}+Q^{3}\right) }.\label{26}
\end{equation}
It is important to emphasize that the Gibbs free energy plays a crucial role in determining the thermodynamic stability of the black hole. Specifically, when $G$ vanishes, the system reaches a critical threshold that separates stable and unstable phases. From Eq. (\ref{26}), we find that this critical point occurs at:
\begin{equation}
r_{H}=0.623\times Q.
\end{equation}
At this horizon radius, the Gibbs free energy tends to zero, indicating a potential phase transition in the black hole system.

\begin{figure}[H]
\begin{minipage}[t]{0.33\textwidth}
        \centering
            \includegraphics[width=\textwidth]{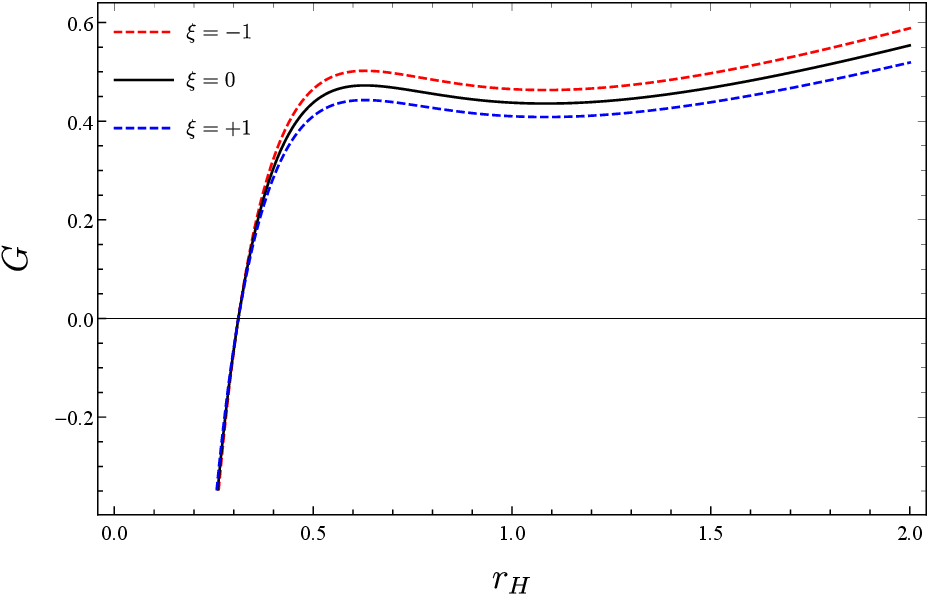}
            \subcaption{$\eta=0.05$}
         \label{fig:g1}
\end{minipage}%
\begin{minipage}[t]{0.33\textwidth}
        \centering
               \includegraphics[width=\textwidth]{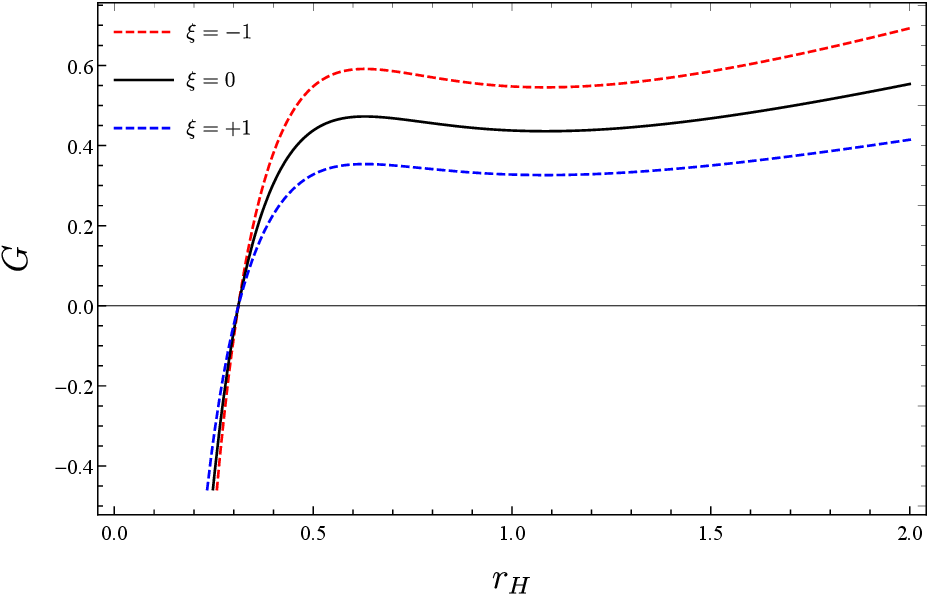}
                \subcaption{$\eta=0.1$}
       \label{fig:g2}
   \end{minipage}%
\begin{minipage}[t]{0.33\textwidth}
        \centering
                \includegraphics[width=\textwidth]{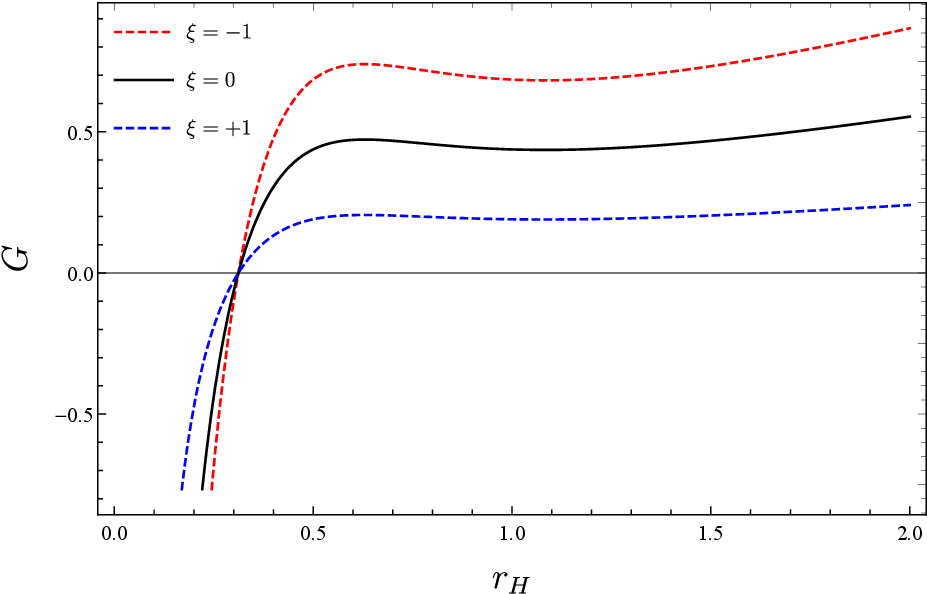}
                \subcaption{$\eta=0.15$}
         \label{fig:g3}
   \end{minipage}\caption{Gibbs free energy $G$ as a function of the event horizon radius $r_H$ for different values of the symmetry-breaking parameter $\eta$. The plots illustrate the existence of a second-order phase transition, where $G$ changes sign at a critical radius $r_H^{c}$. For small black holes, $G>0$ suggests instability, whereas for larger black holes, $G<0$ indicates a stable phase.}
\label{fig:G}
\end{figure}

In Figure \ref{fig:G}, we plot the Gibbs free energy as a function of the event horizon radius $r_H$ for different values of the symmetry-breaking parameter $\eta$. 

\noindent The plots reveal the presence of a second-order phase transition. Notably, for small $r_H$, the Gibbs free energy is positive, suggesting an unstable phase, whereas for larger values of $r_H$, the free energy becomes negative, indicating a thermodynamically stable phase. The presence of the phantom global monopole ($\xi=-1$) shifts the critical point, influencing the black hole's stability conditions.

\section{Timelike Geodesics} \label{sec4}
In GR, test particles move along geodesics, which represent the natural trajectories dictated by the underlying spacetime curvature. When no external forces act upon a particle, its motion is governed solely by the spacetime geometry \cite{Cardoso:2008bp, Konoplya:2017wot, Guo1, Konoplya:2022gjp}. The nature of a geodesic depends on the mass of the particle: massive particles traverse timelike geodesics, whereas massless particles, such as photons, follow null geodesics. The geodesic equation, which describes these trajectories, is given by  
\begin{equation}
\frac{d^{2}x^{\nu }}{d\lambda ^{2}}+\Gamma _{\mu \sigma }^{\nu }\frac{dx^{\mu }}{d\lambda }\frac{dx^{\sigma }}{d\lambda }=0,
\end{equation}
where \( \lambda \) represents an affine parameter along the geodesic, and \( \Gamma _{\mu \sigma }^{\nu } \) are the Christoffel symbols associated with the spacetime metric. 

For a spherically symmetric black hole spacetime, the geodesic motion simplifies, leading to the following system of differential equations:
\begin{equation}
\ddot{t}+\frac{f^{\prime }(r)}{f(r)} \dot{r} \dot{t} = 0,
\end{equation}
\begin{equation}
\ddot{r}+\frac{f(r)}{2} \left( f^{\prime }(r) \dot{t}^{2} + \frac{\dot{r}}{f^{\prime }(r)} - 2r\dot{\theta}^{2} - 2r\sin^2\theta \dot{\phi}^{2} \right) = 0,
\end{equation}
\begin{equation}
\ddot{\theta} + 2\frac{\dot{r}}{r} \dot{\theta} - \cos\theta \sin\theta \dot{\phi}^{2} = 0,
\end{equation}
\begin{equation}
\ddot{\phi} + 2\frac{\dot{r}}{r} \dot{\phi} + 2\cot\theta \dot{\theta} \dot{\phi} = 0.
\end{equation}
A notable property of geodesic motion in spherically symmetric spacetimes is that if a particle initially moves within the equatorial plane ($\theta = \pi/2$ at an initial time $t_i$), it remains confined to this plane throughout its motion. This follows directly from the equation for $\ddot{\theta}$, which vanishes under these conditions. As a result, the motion can be analyzed without loss of generality by restricting to the equatorial plane, leading to the simplified geodesic equations:
\begin{equation}
\ddot{t}+\frac{f^{\prime }(r)}{f(r)} \dot{r} \dot{t} = 0,
\end{equation}
\begin{equation}
\ddot{r}+\frac{f(r)}{2} \left( f^{\prime }(r) \dot{t}^{2} + \frac{\dot{r}}{f^{\prime }(r)} - 2r\dot{\phi}^{2} \right) = 0,
\end{equation}
\begin{equation}
\ddot{\phi} + 2\frac{\dot{r}}{r} \dot{\phi} = 0.
\end{equation}
Geodesic motion obeys a conserved quantity that arises due to metric compatibility:
\begin{equation}
g_{\mu \nu } \frac{dx^{\mu }}{d\lambda } \frac{dx^{\nu }}{d\lambda } = -1,
\end{equation}
Due to the presence of Killing vector fields associated with the metric symmetries, the geodesic equations admit two conserved quantities: the specific energy $E$ and the specific angular momentum $L$. These conserved quantities simplify the equations of motion:
\begin{equation}
\dot{t} = \frac{E}{f(r)},\label{11}
\end{equation}
\begin{equation}
\dot{\phi} = \frac{L}{r^{2}}.\label{ph}
\end{equation}
By expressing the radial motion in terms of these conserved quantities, we obtain a first-order differential equation:
\begin{equation}
\dot{r}^{2} + V_{\mathrm{eff}}(r) = E^{2},  \label{eqv}
\end{equation}
where the effective potential governing the radial motion is given by:
\begin{equation}
V_{\mathrm{eff}}(r) = \left( 1 - 8\pi \xi \eta^{2} - \frac{2mr^{2}}{r^{3} + Q^{3}} \right) \left( 1 + \frac{L^{2}}{r^{2}} \right).
\end{equation}
Eq. (\ref{eqv}) describes a system analogous to the motion of a classical particle with energy $ E^2 $ in an effective one-dimensional potential $ V_{\mathrm{eff}}(r) $. The nature of this potential determines key features of the geodesic motion, such as stable circular orbits, escape conditions, and critical impact parameters for light bending. The influence of the phantom field, NLE, and the global monopole manifests in modifications to this potential, altering the behavior of both massive and massless test particles.

\begin{figure}[H]
\begin{minipage}[t]{0.33\textwidth}
        \centering
            \includegraphics[width=\textwidth]{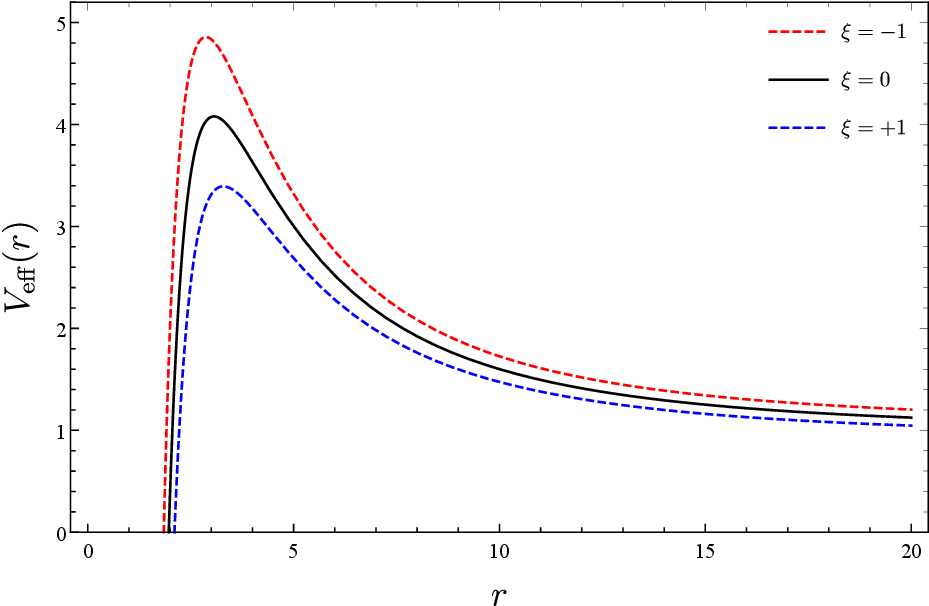}
            \subcaption{$\eta=0.05$}
         \label{fig:ve1}
\end{minipage}%
\begin{minipage}[t]{0.33\textwidth}
        \centering
               \includegraphics[width=\textwidth]{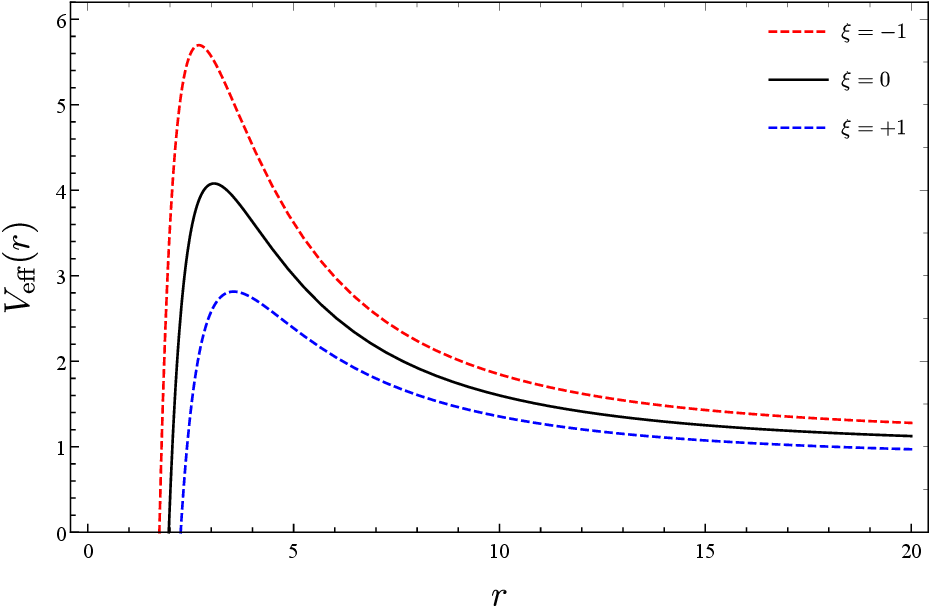}
                \subcaption{$\eta=0.07$}
       \label{fig:ve2}
   \end{minipage}%
\begin{minipage}[t]{0.33\textwidth}
        \centering
                \includegraphics[width=\textwidth]{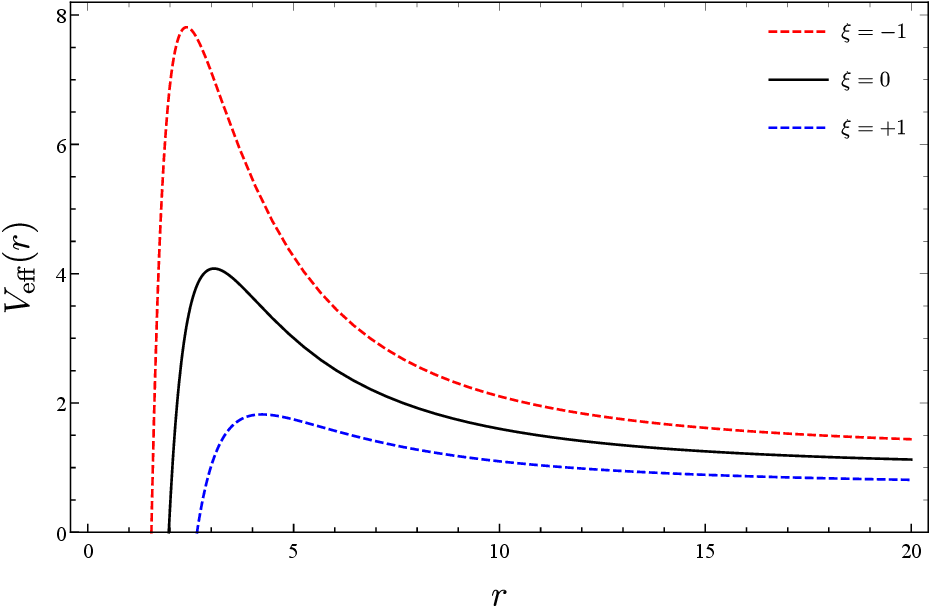}
                \subcaption{$\eta=0.1$}
         \label{fig:ve3}
   \end{minipage}
\caption{Variation of the effective potential $V_{\mathrm{eff}}(r)$ with different values of $\eta$, considering $Q =0.5$ and $m =1$.}
\label{veeff1}
\end{figure}

On the other hand, for the motion of massive test particles, the effective potential plays a crucial role in determining orbital dynamics. It governs the radial motion and establishes the conditions for stable orbits. The effective potential for a test particle in this spacetime is given by:

\begin{equation}
V_{\mathrm{eff}}(r) = \left( 1 - \frac{2mr^{2}}{r^{3} + Q^{3}} \right) + \frac{L^{2}}{r^{2}} - \frac{2m}{r^{3} + Q^{3}} L^{2} - 8\pi \xi \eta^{2} \left( 1 + \frac{L^{2}}{r^{2}} \right).
\label{14}
\end{equation}
The terms in this expression contribute distinctively to the potential landscape: the second term represents a centrifugal barrier, the third term introduces relativistic corrections due to spacetime curvature, and the fourth term accounts for modifications arising from the presence of the phantom global monopole. Figure \ref{veeff1} illustrates the behavior of $V_{\mathrm{eff}}(r)$ for varying values of $\eta$. 

Analyzing Eq. (\ref{14}) asymptotically, we observe that at large distances, the effective potential approaches a limiting value:
\begin{equation}
V_{\mathrm{eff}}(r) \rightarrow \left( 1 - 8\pi \xi \eta^{2} \right) \quad \text{as} \quad r \to \infty.
\end{equation}
For a particle to escape to infinity, its total energy must satisfy the condition:
\begin{equation}
E^{2} \geq 1 - 8\pi \xi \eta^{2}.
\end{equation}
Orbits that satisfy this condition are classified as unbound. In contrast, bound orbits require:
\begin{equation}
E^{2} \leq 1 - 8\pi \xi \eta^{2}.
\end{equation}
Here, the term $1 - 8\pi \xi \eta^{2}$ represents the effective rest energy of the test particle in this modified spacetime.

\begin{figure}[H]
\begin{minipage}[t]{0.33\textwidth}
        \centering
            \includegraphics[width=\textwidth]{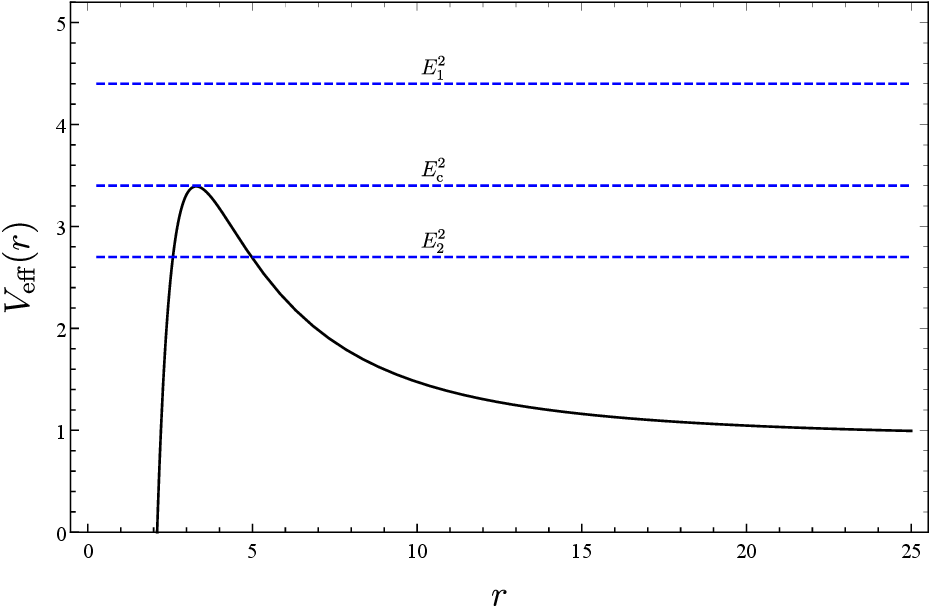}
            \subcaption{$\xi=1$}
         \label{fig:v1}
\end{minipage}%
\begin{minipage}[t]{0.33\textwidth}
        \centering
               \includegraphics[width=\textwidth]{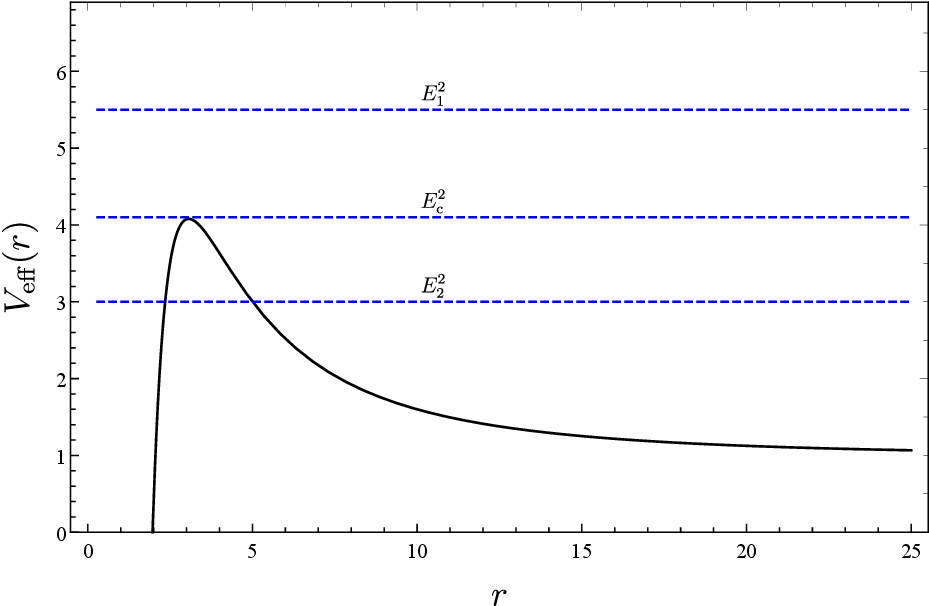}
                \subcaption{$\xi=0$}
       \label{fig:v2}
   \end{minipage}%
\begin{minipage}[t]{0.33\textwidth}
        \centering
                \includegraphics[width=\textwidth]{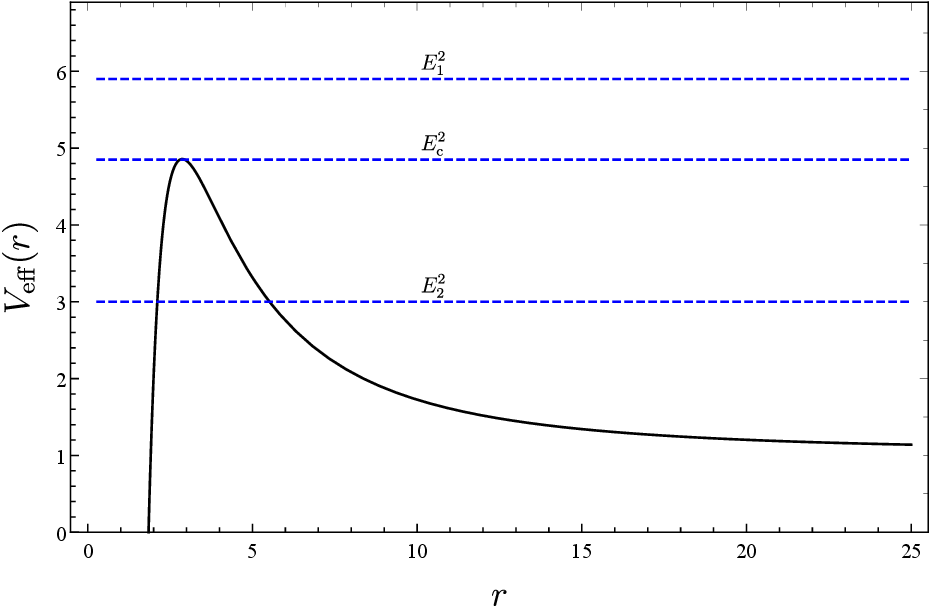}
                \subcaption{$\xi=-1$}
         \label{fig:v3}
   \end{minipage}
\caption{Effective potential for different values of $\xi$, with fixed $Q =0.5$, $\eta =0.05$, $L=10$, and $m =1$.}
\label{veff1}
\end{figure}

To explore the nature of these orbits further, one can categorize the particle’s motion based on its energy relative to a critical value $E_c$:
\begin{itemize}
    \item $E > E_c$: The particle has enough energy to escape or fall directly into the singularity.
    \item $E = E_c$: The particle undergoes unstable circular motion.
    \item $E < E_c$: The particle follows a bound orbit, approaching a minimum radius before returning outward.
\end{itemize}
These behaviors are depicted in Figure \ref{veff1}, illustrating the variation of the effective potential under different values of the parameter $\xi$.

\begin{figure}[H]
\begin{minipage}[t]{0.33\textwidth}
        \centering
            \includegraphics[width=\textwidth]{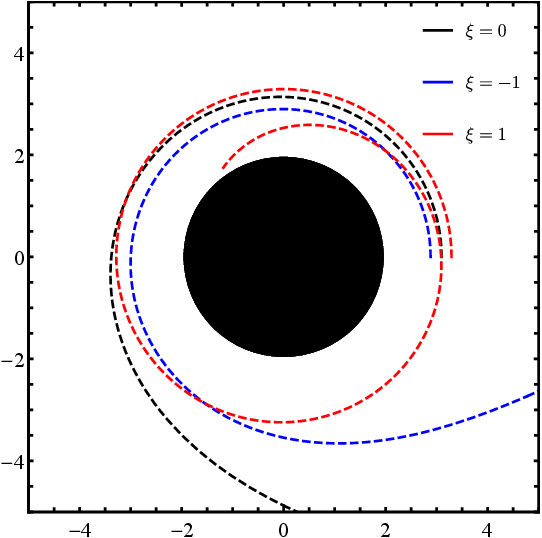}
            \subcaption{$L=10$}
         \label{fig:t1}
\end{minipage}%
\begin{minipage}[t]{0.33\textwidth}
        \centering
               \includegraphics[width=\textwidth]{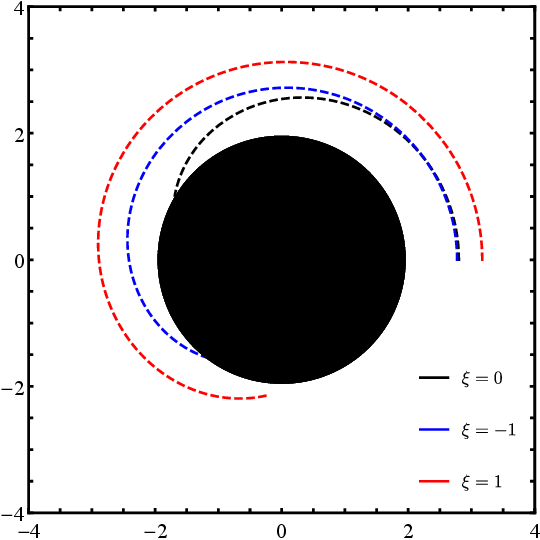}
                \subcaption{$L=20$}
       \label{fig:t2}
   \end{minipage}%
\begin{minipage}[t]{0.33\textwidth}
        \centering
                \includegraphics[width=\textwidth]{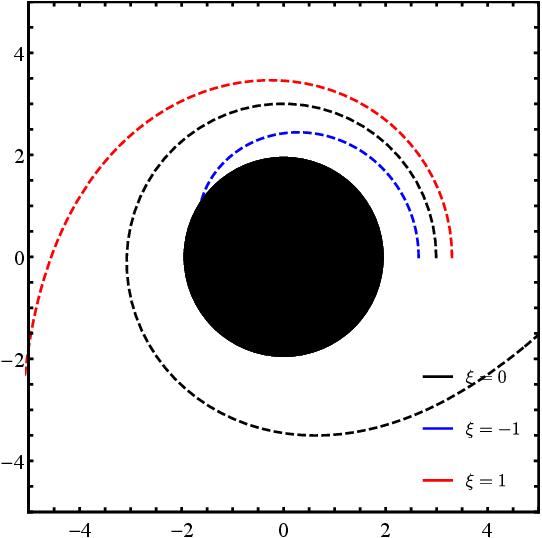}
                \subcaption{$L=30$}
         \label{fig:t3}
   \end{minipage}
\caption{Unstable circular orbits for different angular momentum values, considering $Q =0.5$, $\eta =0.05$, and $m =1$.}
\label{trajectoire}
\end{figure}

\noindent Furthermore, Figure \ref{trajectoire} presents the trajectories of test particles in unstable circular orbits for varying angular momentum values.

\begin{figure}[H]
\begin{minipage}[t]{0.33\textwidth}
        \centering
            \includegraphics[width=\textwidth]{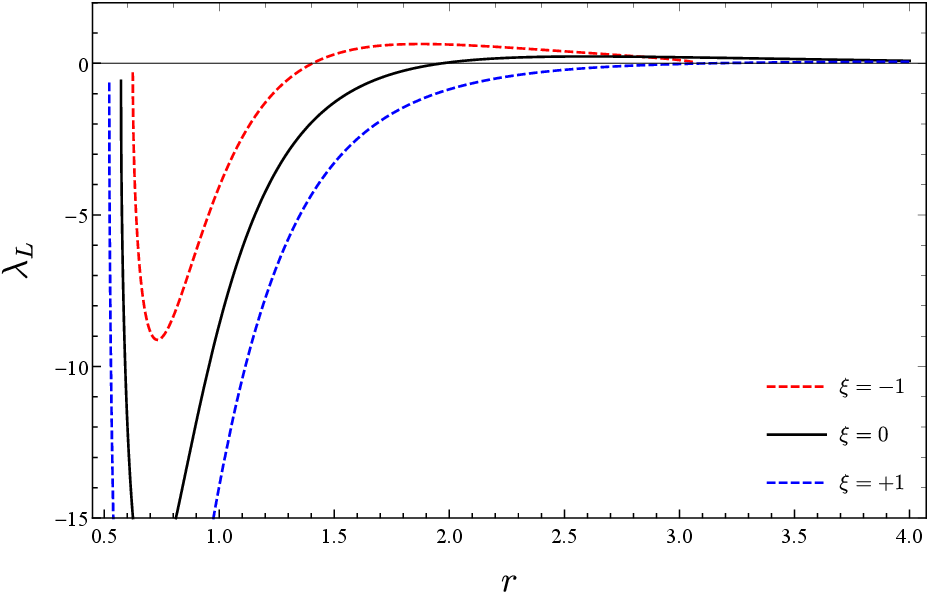}
            \subcaption{$\eta=0.12$}
         \label{fig:L1}
\end{minipage}%
\begin{minipage}[t]{0.33\textwidth}
        \centering
               \includegraphics[width=\textwidth]{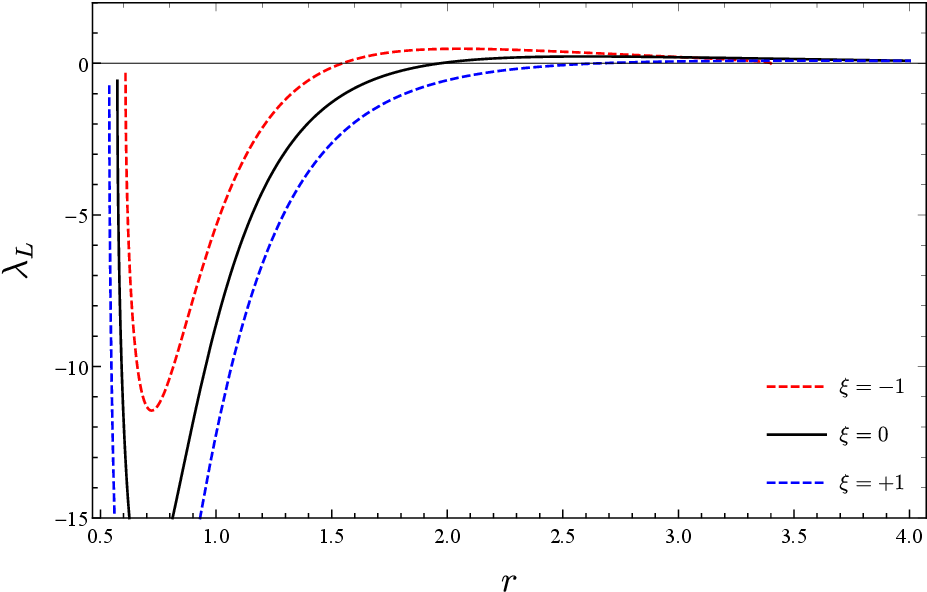}
                \subcaption{$\eta=0.1$}
       \label{fig:L2}
   \end{minipage}%
\begin{minipage}[t]{0.33\textwidth}
        \centering
                \includegraphics[width=\textwidth]{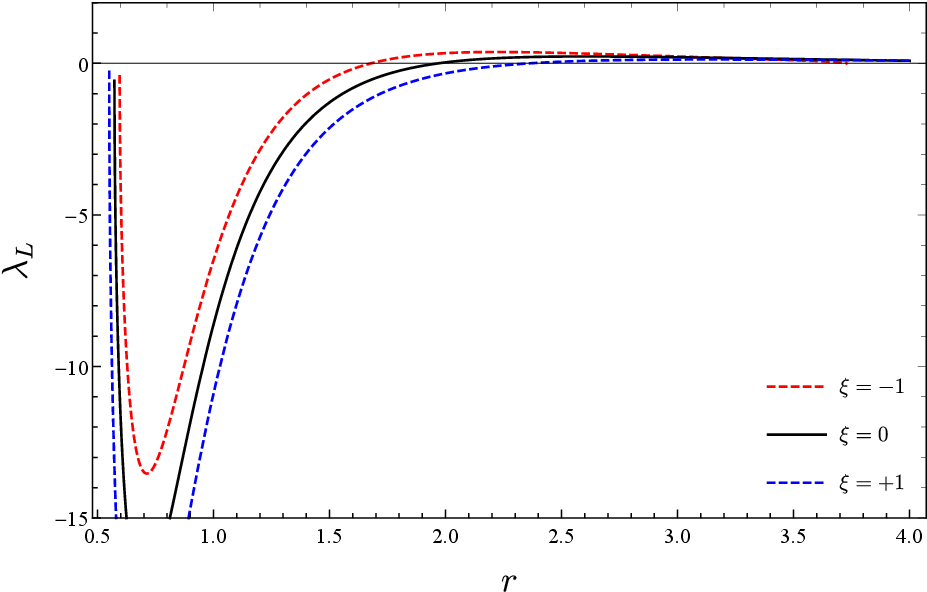}
                \subcaption{$\eta=0.08$}
         \label{fig:L3}
   \end{minipage}
\caption{Radial dependence of the Lyapunov exponent $\lambda_L$ for test particle motion around a nonlinear magnetic-charged black hole with a phantom global monopole. The parameters used are $Q =0.5$, $L =5$, $E =1$, and $m =1$.}
\label{Lyapunov}
\end{figure}

\noindent For a qualitative analysis of circular orbits, it is necessary to examine the conditions under which the radial coordinate remains fixed. These conditions are derived by setting:
\begin{equation}
\dot{r} = 0 \quad \Longrightarrow \quad V_{\mathrm{eff}}(r) = E^{2},\label{16}
\end{equation}
\begin{equation}
\ddot{r} = 0 \quad \Longrightarrow \quad V_{\mathrm{eff}}^{\prime}(r) = 0,\label{17}
\end{equation}
where $V_{\mathrm{eff}}^{\prime}(r) = \frac{\partial V_{\mathrm{eff}}(r)}{\partial r}$. Applying these conditions allows us to determine the specific energy $E$ and angular momentum $L$ required for stable circular orbits:
\begin{equation}
L^{2}=\frac{r^{7}-2Q^{3}r^{4}}{Q^{6}-8\pi \eta ^{2}\xi \left(Q^{3}+r^{3}\right) ^{2}+2Q^{3}r^{3}+r^{6}-3r^{5}},
\end{equation}%
\begin{equation}
E^{2}=\frac{4\left( r^{4}-2Q^{3}r\right) ^{2}}{\left( Q^{3}+r^{3}\right)^{2}\left( Q^{6}+2Q^{3}r^{3}+(r-3)r^{5}\right)
-8\pi \eta ^{2}\xi \left(Q^{3}+r^{3}\right) ^{4}}.
\end{equation}
The stability of these orbits is determined by the second derivative of the effective potential:
\begin{eqnarray}
 \frac{\partial^{2} V_{\mathrm{eff}}(r) }{\partial r^{2}}>0 \quad &\Rightarrow& \quad \text{Unstable}, \\ 
 \frac{\partial^{2} V_{\mathrm{eff}}(r) }{\partial r^{2}}<0 \quad &\Rightarrow& \quad \text{Stable}.
\end{eqnarray}
The innermost stable circular orbit (ISCO) is found by imposing the additional condition:
\begin{equation}
\frac{\partial^{2} V_{\mathrm{eff}}(r) }{\partial r^{2}} = 0.
\end{equation}
A deeper understanding of orbital motion can be gained by analyzing circular orbits, where the radial coordinate $r$ remains constant. To sustain a circular trajectory, a test particle must satisfy two conditions. The first ensures that the radial velocity vanishes:
\begin{equation}
\dot{r} = 0 \quad \Longrightarrow \quad V_{\mathrm{eff}}(r) = E^{2}.
\label{16a}
\end{equation}
The second condition ensures that there is no radial acceleration:
\begin{equation}
\ddot{r} = 0 \quad \Longrightarrow \quad V_{\mathrm{eff}}^{\prime}(r) = 0,
\label{17a}
\end{equation}
where the prime denotes differentiation with respect to $r$. Using these conditions, one can determine the specific angular momentum $L$ and specific energy $E$ required for a test particle to maintain a circular orbit:
\begin{equation}
L^{2} = \frac{r^{7} - 2Q^{3}r^{4}}{Q^{6} - 8\pi \eta^{2} \xi (Q^{3} + r^{3})^{2} + 2Q^{3}r^{3} + r^{6} - 3r^{5}},
\end{equation}
\begin{equation}
E^{2} = \frac{4(r^{4} - 2Q^{3}r)^{2}}{(Q^{3} + r^{3})^{2} (Q^{6} + 2Q^{3}r^{3} + (r-3)r^{5}) - 8\pi \eta^{2} \xi (Q^{3} + r^{3})^{4}}.
\end{equation}
The stability of these circular orbits is determined by the second derivative of the effective potential:
\begin{eqnarray}
\frac{\partial^{2} V_{\mathrm{eff}}(r) }{\partial r^{2}}>0 \quad &\Rightarrow& \quad \text{Unstable}, \\ 
\frac{\partial^{2} V_{\mathrm{eff}}(r) }{\partial r^{2}}<0 \quad &\Rightarrow& \quad \text{Stable}.
\end{eqnarray}
The innermost stable circular orbit (ISCO) is of particular significance as it marks the transition between stable and unstable motion. At the ISCO, the maximum and minimum of the effective potential merge, imposing the additional condition:
\begin{equation}
\frac{\partial^{2} V_{\mathrm{eff}}(r) }{\partial r^{2}} = 0.
\end{equation}
By substituting Eq. (\ref{14}) into this condition, we derive the ISCO constraint:
\begin{equation}
\left. rf^{\prime \prime }(r) f(r) +3f(r) f^{\prime }(r) -2rf^{\prime }(r)^{2}\right|_{r = r_{\mathrm{ISCO}}} = 0.
\end{equation}
Expanding this equation explicitly, we obtain:
\begin{equation}
\left. 11Q^{3}r^{4}(1 - 8\pi \eta^{2} \xi) - 8Q^{6}r(1 - 8\pi \eta^{2} \xi) + r^{6} (r - 6 - 8\pi \eta^{2} \xi r) \right|_{r = r_{\mathrm{ISCO}}} = 0.
\end{equation}
To gain further insight into the ISCO properties, we numerically evaluate the ISCO radius $r_{\mathrm{ISCO}}$, as well as the corresponding angular momentum $L_{\mathrm{ISCO}}$ and energy $E_{\mathrm{ISCO}}$, for different values of $\eta$ and $\xi$. The results are summarized in Table \ref{tabisco}.
\begin{table}[H]
\centering
\begin{tabular}{|l|lll|lll|}
\hline\hline
\rowcolor{lightgray} \multirow{2}{*}{} & \multicolumn{3}{|c|}{$\xi =1$} & 
\multicolumn{3}{|c|}{$\xi =-1$} \\ \hline
\rowcolor{lightgray} $\eta $ & $r_{\mathrm{ISCO}}$ & $L_{\mathrm{ISCO}}$ & $E_{\mathrm{ISCO}}$ & $%
r_{\mathrm{ISCO}}$ & $L_{\mathrm{ISCO}}$ & $E_{\mathrm{ISCO}}$ \\ \hline
$0.01$ & 5.97664 & 3.46678 & 0.0793058 & 5.94608 & 3.44931 & 0.0799223 \\ 
$0.05$ & 6.36837 & 3.69102 & 0.0720584 & 5.6015 & 3.25243 & 0.0874702 \\ 
$0.09$ & 7.50929 & 4.34573 & 0.0562133 & 4.92859 & 2.8693 & 0.106192 \\ 
$0.12$ & 9.38748 & 5.42642 & 0.0401827 & 4.3324 & 2.5321 & 0.129227 \\ 
\hline\hline
\end{tabular}
\label{tabs}
\caption{Numerical values of the ISCO parameters $r_{\mathrm{ISCO}}$, $L_{\mathrm{ISCO}}$, and $E_{\mathrm{ISCO}}$ for test particles under different values of $\eta$ and $\xi$.}
\label{tabisco}
\end{table}
Another key aspect of orbital stability that we analyze is the principal Lyapunov exponent, which characterizes the divergence or convergence of neighboring trajectories in phase space. This exponent quantifies how small perturbations affect the stability of circular orbits. A positive Lyapunov exponent indicates an unstable orbit where nearby trajectories diverge over time, whereas a negative Lyapunov exponent corresponds to a stable orbit with converging trajectories. The Lyapunov exponent $\lambda_{L}$ is defined as:
\begin{equation}
\lambda_{L} = \sqrt{-\frac{1}{2\dot{t}^{2}} \frac{\partial^{2} V_{\mathrm{eff}}(r)}{\partial r^{2}}}.
\end{equation}
By substituting Eqs. (\ref{11}) and (\ref{14}) into the above expression, we derive the explicit form of $\lambda_L$ for a nonlinear magnetic-charged black hole with a phantom global monopole:
\begin{eqnarray}
\lambda_{L} &=& \frac{1}{E} \left( 1 - 8\pi \eta^{2} \xi - \frac{2m r^{2}}{Q^{3} + r^{3}} \right)  
\Biggr[\frac{2L^{2}}{r^{3}} \left( \frac{6m r^{4}}{(Q^{3} + r^{3})^{2}} - \frac{4m r}{Q^{3} + r^{3}} \right) - \frac{3L^{2}}{r^{4}} 
\left( 1 - 8\pi \eta^{2} \xi - \frac{2m r^{2}}{Q^{3} + r^{3}} \right) \notag \\
&& + \frac{m}{Q^{3} + r^{3}} \left( 1 + \frac{L^{2}}{r^{2}} \right) \left( 1 - \frac{9r^{3}}{(Q^{3} + r^{3})} + \frac{9r^{6}}
{(Q^{3} + r^{3})^{2}} \right) \Biggr]^{1/2}.
\end{eqnarray}
In Figure \ref{Lyapunov}, we illustrate the radial dependence of the Lyapunov exponent for test particles orbiting a nonlinear magnetic-charged black hole with a phantom global monopole. The behavior of $\lambda_L$ provides crucial insight into the stability of circular orbits. 

\noindent From Figure \ref{Lyapunov}, we observe that as the energy scale of symmetry breaking, represented by $\eta$, decreases, the region allowing stable circular orbits expands. This suggests that a weaker global monopole field enhances orbital stability by reducing the disruptive effects on geodesic motion. Additionally, the Lyapunov exponent curves intersect the horizontal axis at specific radii, marking the location of the ISCO. These bifurcation points indicate a transition between stable and unstable circular orbits. Furthermore, as $\eta$ decreases, the ISCO radius shifts outward, implying that particles in lower-energy monopole fields require a larger orbital radius to maintain stability. This trend highlights the significant role of the global monopole in shaping the phase-space dynamics of test particle motion in modified gravity scenarios.

\section{Quasinormal Modes} 
\label{sec5}

QNMs characterize the response of a black hole to external perturbations and provide crucial insights into its stability and observational signatures in gravitational wave astronomy \cite{Kokkotas1999, Konoplya:2011qq}. The spectrum of these modes depends on the black hole parameters and surrounding matter fields, making them valuable probes of deviations from general relativity. 

In the presence of a global monopole, the quasinormal spectrum is modified due to the conical deficit introduced by the monopole field. Prior studies have examined quasinormal ringing in such backgrounds for scalar \cite{Zhang2005, Chang2006, Zhou2013a}, Dirac \cite{Zhou2013b}, Maxwell \cite{Wang2021}, and spinor fields \cite{MoraisGraca:2015snz}. These works show that the presence of a monopole shifts the real and imaginary parts of the QNM frequencies, affecting the decay rate of perturbations. More recent research in modified gravity scenarios, such as Einstein–bumblebee gravity, has further explored the influence of global monopoles on black hole stability and QNM spectra \cite{Lin2023, Zhang2023, Malik:2023bxc}.  

Additionally, NLE introduces further modifications to the QNM spectrum by altering the effective potential governing perturbations. These effects have been investigated in contexts such as Einstein-Gauss-Bonnet gravity \cite{Kruglov2021}, AdS black holes with Coulomb-like corrections \cite{González2021}, and charged black holes with NLE-induced modifications \cite{Nomura2022, Sing2022}. Studies have also examined how these effects influence black hole shadows and grey-body factors \cite{Okyay2022, Wu2024}, while additional work on Euler-Heisenberg black holes suggests notable shifts in the QNM spectrum \cite{Hamil2025fort}.  

In this section, we analyze the QNMs of a nonlinear magnetic-charged black hole with a phantom global monopole. Using the sixth-order WKB approximation, we compute the mode spectrum and compare our results with the Pöschl-Teller potential approach to assess accuracy. Our goal is to determine how the interplay between the phantom global monopole and NLE influences the black hole’s dynamical response and stability under scalar perturbations.

Let us consider the dynamics of a massless scalar field propagating in the black hole background, governed by the wave equation:
\begin{equation}
\frac{1}{\sqrt{-g}}\partial_{\mu} \left( \sqrt{-g} g^{\mu \nu} \partial_{\nu} \psi \right) = 0.
\end{equation}
Applying the method of separation of variables, we express the field as:
\begin{equation}
\psi_{n,\ell,\mu} (r,\theta,\phi,t) = \frac{R_{n,\ell} (r)}{r} \mathcal{Y}_{\ell,\mu} (\theta,\phi) e^{-i\omega t},
\end{equation}
where $\mathcal{Y}_{\ell,\mu} (\theta,\phi)$ are the spherical harmonics describing the angular dependence, and $R_{n,\ell} (r)$ is the radial function satisfying the equation:
\begin{equation}
f(r) \frac{d}{dr} \left[ f(r) \frac{d}{dr} R_{n,\ell} (r) \right] 
+ \Big( \omega^2 - \mathcal{V}_{\mathrm{eff}} (r) \Big) R_{n,\ell} (r) = 0.
\end{equation}
Here, the effective potential governing the scalar perturbations is given by:
\begin{equation}
\mathcal{V}_{\mathrm{eff}} (r) = f(r) \left( \frac{\ell (\ell +1)}{r^{2}} 
+ \frac{1}{r} \frac{d}{dr} f(r) \right).
\end{equation}
To further simplify the radial equation, we introduce the tortoise coordinate $x$:
\begin{equation}
x = \int \frac{dr}{f(r)},
\end{equation}
which transforms the radial equation into a Schrödinger-like form:
\begin{equation}
\frac{d^{2}}{dx^{2}} R_{n,\ell} (x) + \Big( \omega^2 - \mathcal{V}_{\mathrm{eff}} (x) \Big) R_{n,\ell} (x) = 0.
\label{52}
\end{equation}
The behavior of the QNMs is determined by imposing physically motivated boundary conditions on Eq.~(\ref{52}). These conditions ensure purely ingoing waves at the black hole horizon and purely outgoing waves at spatial infinity, leading to:
\begin{equation}
R_{n,\ell} (x) \thickapprox e^{\pm i \omega x} \quad \text{as} \quad x \longrightarrow \pm \infty.
\label{condi}
\end{equation}
The effective potential $\mathcal{V}_{\mathrm{eff}}(x)$ determines the QNM spectrum, as it represents the potential barrier through which perturbations must tunnel before being absorbed by the black hole. Its shape and peak height influence the stability and damping rates of the modes. In the following, we analyze the numerical methods used to compute the QNM spectrum and discuss the effects of the phantom global monopole and NLE on the stability of the black hole.

\subsection{Computation of QNMs using the WKB Approximation}

To compute the quasinormal frequencies, we employ the WKB approximation, a widely used semi-analytical method in black hole physics \cite{Konoplya:2006ar, Kodama:2009bf, Konoplya:2001ji, Konoplya:2005sy, Konoplya:2023ahd, Ham1, Hamil20241, Hamil2025fort, Konoplya20251, Lutfuoglu2025, Chen2022, Marcos2021, Marcos2022, Lutfuoglu:2025ljm, Lutfuoglu:2025hwh}. Originally introduced at first order in \cite{Schutz:1985km}, the method was later extended to higher orders, including second and third order in \cite{Iyer:1986np} and up to sixth order in \cite{Konoplya:2003ii}. While higher-order corrections enhance precision, their effectiveness depends on the potential shape and parameter regime, as discussed in the reviews \cite{Konoplya:2011qq, Konoplya:2019hlu}.

In this work, we use the sixth-order WKB formula from \cite{Konoplya:2003ii}, given by:
\begin{equation}
\frac{i\left( \omega_{n}^{2}-V_{0}\right) }{\sqrt{-2V_{0}^{\prime \prime }}}
-\sum_{j=2}^{6}\Phi_{j}=n+\frac{1}{2}, \label{53}
\end{equation}
where $V_{0}$ represents the peak of the effective potential at $x_{0}$, the notation ($^{\prime \prime }$) denotes the second derivative with respect to the tortoise coordinate, $\Phi_{j}$ accounts for higher-order corrections, and $n$ is the overtone number. The resulting quasinormal frequencies $\omega_n$ form a discrete spectrum, where the real part corresponds to the oscillation frequency, and the imaginary part determines the damping rate.

Since Eq.~(\ref{53}) involves nontrivial dependencies on physical parameters, we numerically compute the QNM frequencies for different parameter values. However, as the WKB approximation is most reliable when $\ell > n$, we restrict our analysis to scalar field perturbations satisfying this condition, corresponding to the low-lying QNMs. Selected results for these modes are presented in Table~\ref{tab4}, 
\begin{table}[H]
\centering%
\begin{tabular}{|l|l|l|l|l|l|}
\hline\hline
\rowcolor{lightgray} $\ell $ & $n$ & $\omega _{WKB}(\eta =0.12)$ & $\omega
_{WKB}(\eta =0.09)$ & $\omega _{WKB}(\eta =0.06)$ & $\omega _{WKB}(\eta
=0.03)$ \\ \hline
1 & 0 & 0.145742-i 0.0394710 & 0.205796-i 0.0614642 & 0.253484-i 0.0800628 & 0.283985-i 0.0923490 \\ \hline
2 & 0 & 0.244576-i 0.0392214 & 0.342822-i 0.0610029 & 0.420085-i 0.0794028 & 0.469184-i 0.0915496 \\
& 1 & 0.237964-i 0.119079 & 0.331583-i 0.185690 & 0.404717-i 0.242115 & 0.451016-i 0.279432 \\ \hline
3 & 0 & 0.343058-i 0.0391535 & 0.479908-i 0.0608754 & 0.587231-i 0.0792188 &0.655308-i 0.0913261 \\
& 1 & 0.338235-i 0.118187 & 0.471645-i 0.184011 & 0.575871-i 0.239683 &0.641833-i 0.276461 \\  
& 2 & 0.329068-i 0.199332 & 0.456143-i 0.311136 & 0.554751-i 0.405941 &0.616916-i 0.468663 \\ \hline
4 & 0 & 0.441421-i 0.0391258 & 0.617006-i 0.0608231 & 0.754549-i 0.0791432 &0.841728-i 0.0912343 \\  
& 1 & 0.437638-i 0.117818 & 0.610504-i 0.183311 & 0.745590-i 0.238664 &0.831087-i 0.275217 \\ 
& 2 & 0.430296-i 0.197808 & 0.597982-i 0.308271 & 0.728427-i 0.401792 &0.810766-i 0.463613 \\
& 3 & 0.419844-i 0.279875 & 0.580390-i 0.437128 & 0.704537-i 0.570554 & 0.782633-i 0.658860 \\ \hline\hline
\end{tabular}%
\caption{QNMs of the nonlinear magnetic-charge BH with ordinary global monopole ($\xi=1$), calculated using the 6th-order WKB approximation method, for $m=1$ and $Q = 0.5$.}
\label{tab4}
\end{table}

\noindent and in Table \ref{tab5}.

\begin{table}[H]
\centering%
\begin{tabular}{|l|l|l|l|l|l|}
\hline\hline
\rowcolor{lightgray} $\ell $ & $n$ & $\omega _{WKB}(\eta =0.12)$ & $\omega
_{WKB}(\eta =0.09)$ & $\omega _{WKB}(\eta =0.06)$ & $\omega _{WKB}(\eta
=0.03)$ \\ \hline
1 & 0 & 0.483648-i 0.176601 & 0.395847-i 0.139083 & 0.337947-i 0.114637 & 
0.305100-i 0.100997 \\ \hline
2 & 0 & 0.785554-i 0.174813 & 0.647424-i 0.137734 & 0.555511-i 0.113578 & 
0.503043-i 0.100098 \\ 
& 1 & 0.747231-i 0.535822 & 0.618247-i 0.421576 & 0.532151-i 0.347187 & 
0.482875-i 0.305711 \\ \hline
3 & 0 & 1.09191-i 0.174313 & 0.901696-i 0.137354 & 0.774781-i 0.113280 & 
0.702201-i 0.0998456 \\ 
& 1 & 1.06303-i 0.528917 & 0.879837-i 0.416441 & 0.757366-i 0.343204 & 
0.687212-i 0.302355 \\ 
& 2 & 1.01100-i 0.900111 & 0.840080-i 0.707800 & 0.725436-i 0.582622 & 0.659590-i
0.512855 \\ \hline
4 & 0 & 1.39975-i 0.174109 & 1.15685-i 0.137198 & 0.994608-i 0.113158 & 
0.901750-i 0.0997421 \\ 
& 1 & 1.37678-i 0.525991 & 1.13952-i 0.414273 & 0.980824-i 0.341528 & 
0.889903-i 0.300945 \\ 
& 2 & 1.33358-i 0.888399 & 1.10672-i 0.699087 & 0.954632-i 0.575858 & 
0.867323-i 0.507150 \\ 
& 3 & 1.27527-i 1.26657 & 1.06207-i 0.995671 & 0.918684-i 0.819346 & 
0.836173-i 0.721089  \\ \hline\hline
\end{tabular}%
\caption{QNMs of the nonlinear magnetic-charge BH with phantom global monopole ($\xi=-1$), calculated using the 6th-order WKB approximation method, for $m=1$ and $Q = 0.5$.}
\label{tab5}
\end{table}

\noindent The numerical results given in Tables~\ref{tab4} and \ref{tab5} underlines the several key results:

\begin{itemize}
    \item The imaginary part of the frequencies is consistently negative, confirming the stability of scalar field perturbations in this black hole background. This stability arises from the positive potential barrier outside the event horizon, which prevents perturbations from growing unbounded.

    \item As $\ell$ increases for a fixed $\eta$, the real part of the frequency also increases. This trend is expected, as larger angular momentum states correspond to higher oscillation frequencies.

    \item For a given overtone number $n$ and angular momentum $\ell$, increasing the symmetry-breaking parameter $\eta$ generally increases the real part of the frequency in the case of a black hole with an ordinary global monopole. However, for a black hole with a phantom global monopole, the opposite trend is observed—the real part of the QNMs decreases with increasing $\eta$. This distinction suggests that the phantom monopole modifies the effective gravitational potential in a way that reduces oscillation frequencies.  
\end{itemize}

\begin{figure}[H]
\begin{minipage}[t]{0.5\textwidth}
        \centering
        \includegraphics[width=\textwidth]{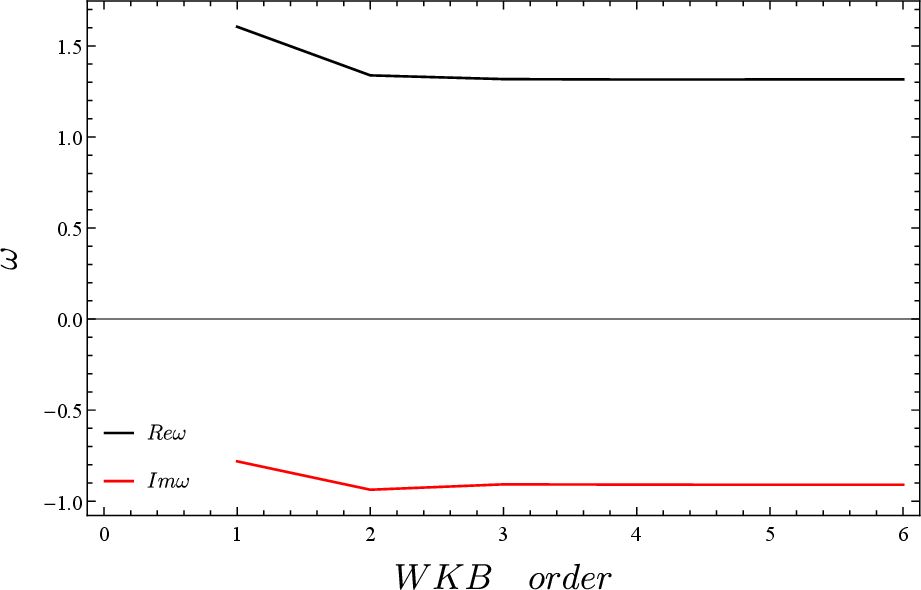}
                \subcaption{$\ell=4$, $n=2$}
        \label{fig:sigma1}
\end{minipage}
\begin{minipage}[t]{0.5\textwidth}
        \centering
        \includegraphics[width=\textwidth]{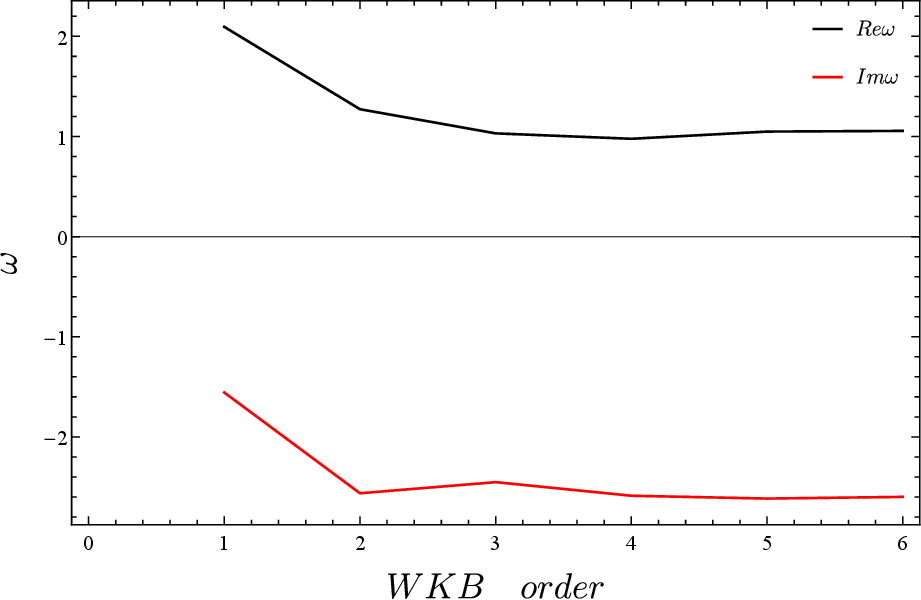}
                \subcaption{$\ell=4$, $n=6$}
        \label{fig:sigma2}
\end{minipage}
\caption{The variation of QNMs for scalar perturbations in a nonlinear magnetic-charged black hole with a phantom global monopole for $\eta = 0.12$, $\xi=-1$ and $Q = 0.3$. The plot illustrates the dependence of QNMs on the overtone number $n$ and angular momentum $\ell$, highlighting the increasing deviation of the WKB approximation for higher overtones. }
\label{fig:sigma}
\end{figure}

\noindent Figure \ref{fig:sigma} illustrates an example of QNMs computed by using the sixth-{order} WKB formula for scalar perturbations with $\eta = 0.12$, $\xi=-1$ and $Q = 0.3$. 

\noindent As evident from the figure, the accuracy of the WKB formula decreases for scalar perturbations when $n \geq {\ell}$, whereas the results converge more reliably for cases where ${\ell} > n$. 

\begin{figure}[H]
\begin{minipage}[t]{0.5\textwidth}
        \centering
        \includegraphics[width=\textwidth]{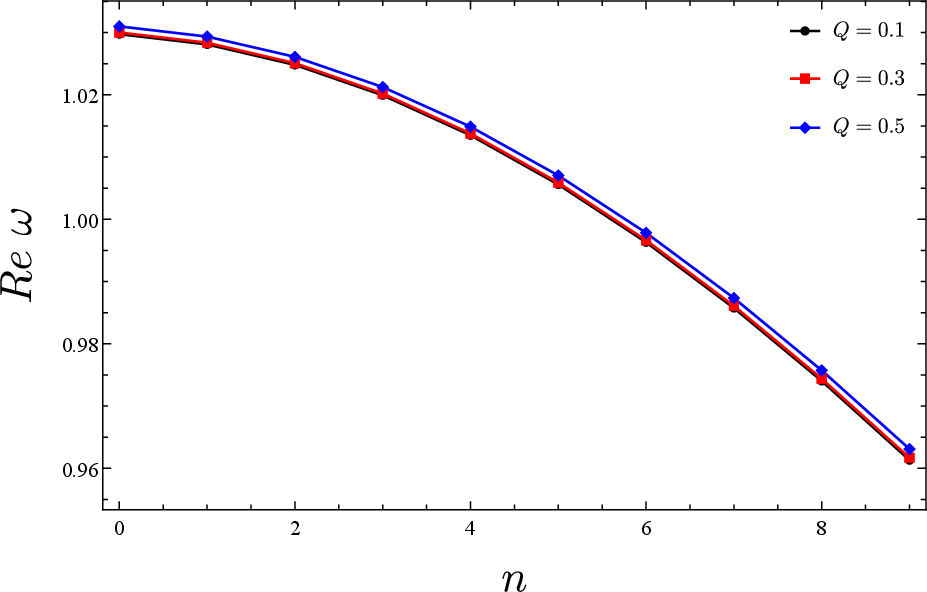}
                \subcaption{$\xi=1$}
        \label{fig:wa}
\end{minipage}
\begin{minipage}[t]{0.5\textwidth}
        \centering
        \includegraphics[width=\textwidth]{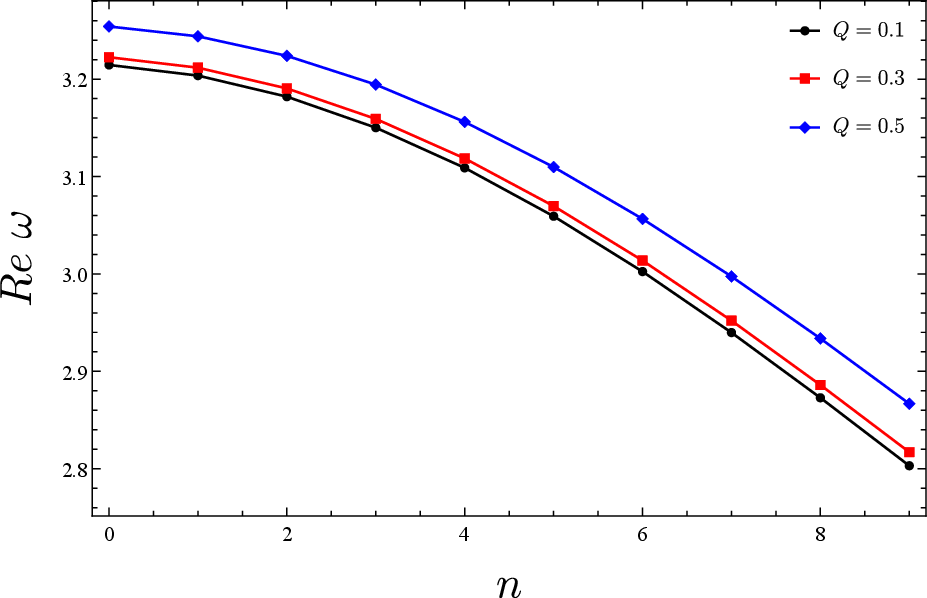}
                \subcaption{$\xi=-1$}
        \label{fig:wb}
\end{minipage}\\
\begin{minipage}[t]{0.5\textwidth}
        \centering
        \includegraphics[width=\textwidth]{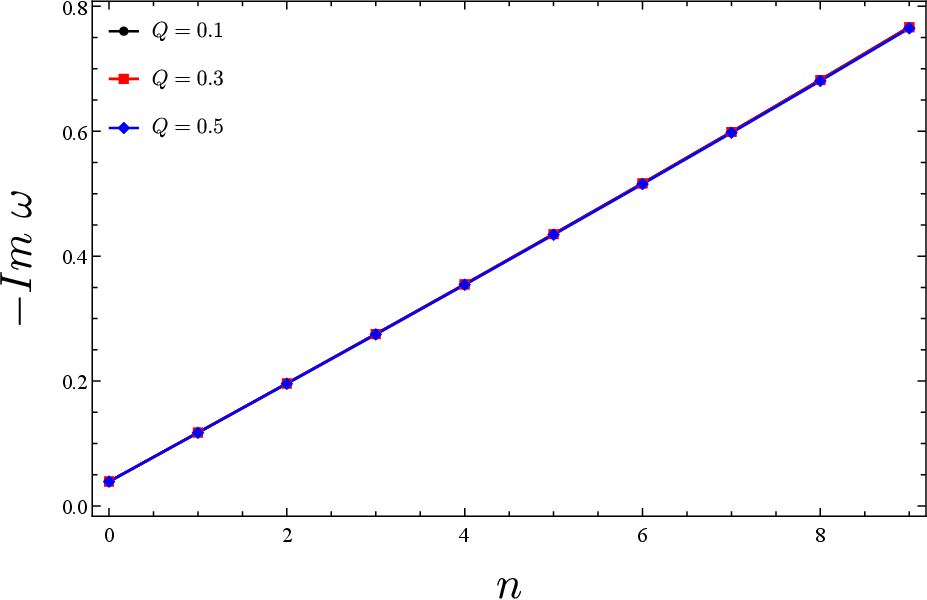}
                \subcaption{$\xi=1$}
        \label{fig:wc}
\end{minipage}
\begin{minipage}[t]{0.5\textwidth}
        \centering
        \includegraphics[width=\textwidth]{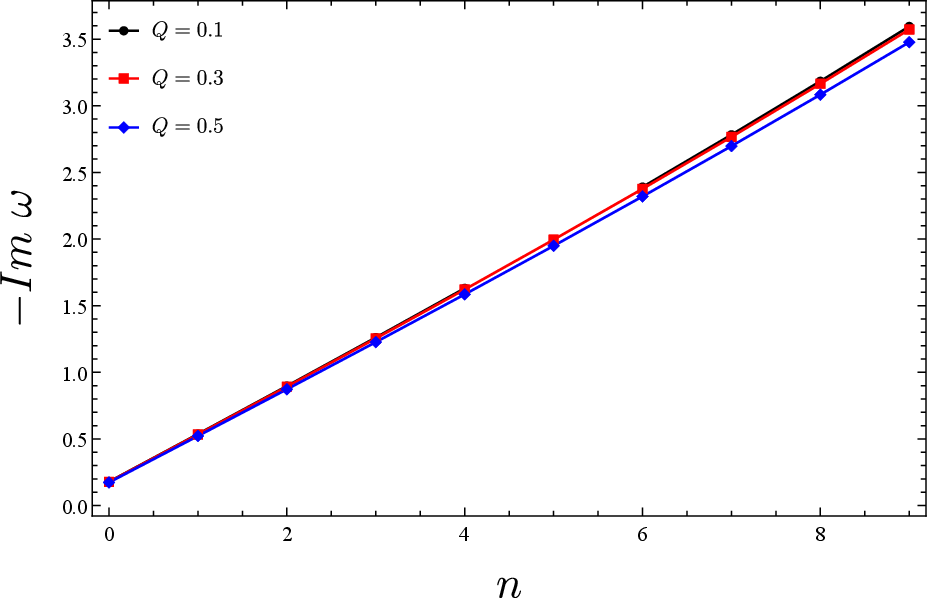}
                \subcaption{$\xi=-1$}
        \label{fig:wd}
\end{minipage}
\caption{QNM frequencies as a function of overtone number $n$ for $\ell=10$, $m=1$, $\eta=0.12$. Higher overtones exhibit lower real frequencies and stronger damping. The effect of the global monopole parameter $\xi$ is also visible.}
\label{fig:QNMn}
\end{figure}

Figure \ref{fig:QNMn} illustrates the changes in the QNMs as a function of $n$ for both ordinary and phantom global monopoles of a nonlinear magnetic-charged black hole. 

\noindent  Figure \ref{fig:QNMn} illustrates the dependence of quasinormal mode frequencies on the overtone number $n$ for scalar perturbations. As $n$ increases, the real part of the frequency decreases, leading to lower oscillation frequencies, while the imaginary part increases, indicating stronger damping. This implies that higher overtones decay more rapidly. Additionally, the presence of a global monopole significantly affects the QNM spectrum, with notable differences between the ordinary ($\xi = 1$) and phantom ($\xi = -1$) cases. The results highlight how the monopole field influences the stability and decay of perturbations in the black hole spacetime.

\subsection{Pöschl-Teller potential approximation}
This approach consists of approximating the potential $\mathcal{V}_{\mathrm{eff}}\left( x\right) $ with a similarly shaped potential, for which the solutions of Eq. (\ref{52}) can be computed analytically. The P\"{o}schl-Teller (PT) potential satisfies this condition and has the following form:%
\begin{equation}
V_{\mathrm{PT}}\left( x\right) =\frac{V_{0}}{\cosh ^{2}\left( \frac{x}{b}%
\right) },
\end{equation}
where
\begin{equation}
 b=\frac{1}{\sqrt{-\frac{1}{2V_{0}}\frac{d^{2}V\left( x_{0}\right) }{dx^{2}}}}.   
\end{equation}
We now adopt the approach outlined in \cite{Mashhoon} to illustrate how the QNM frequencies can be computed analytically for this approximate potential. The key strategy is to transform equation (\ref{52}) into a form that resembles the Schr\"{o}dinger wave equation, such that the solutions of (\ref{52}) that satisfy the boundary condition (\ref{condi}) correspond to bound states of the transformed equation. As is well known, the bound states of the Schr\"{o}dinger equation can be calculated analytically. After computing them, the inverse transformation is applied to obtain the QNMs. The QNMs for the potential $V_{\mathrm{PT}}\left( x\right) $ can be calculated analytically and are expressed as follows:%
\begin{equation}
\omega _{\mathrm{PT}}=\frac{1}{b}\left[- i\left( n+\frac{1}{2}\right) \pm 
\sqrt{V_{0}b^{2}-\frac{1}{4}}\right] .\label{pTeller}
\end{equation}
Now, we compute the quasinormal frequencies for scalar field perturbations for different values of $\eta$, utilizing Eq. (\ref{pTeller}). The results are tabulated in Table \ref{pTellerqnm}. 
\begin{table}[tbh]
\centering%
\begin{tabular}{|l|l|l|l|l|}
\hline\hline
\rowcolor{lightgray} $\eta $ & $\omega _{WKB}(\xi =1)$ & $\omega _{PT}(\xi
=1)$ & $\omega _{WKB}(\xi =-1)$ & $\omega _{PT}(\xi =-1)$ \\ \hline
0.01 & 0.877073 -i 0.0956334 & 0.297450-i 0.100073 & 0.883812 - i 0.0965986  & 
0.299827 -i 0.101102 \\ 
0.02 & 0.866999 -i 0.0941947 & 0.293899-i 0.0985414 & 0.893955 -i 0.0980555
& 0.303406 -i 0.102654 \\ 
0.03 & 0.850299 -i 0.0918209  & 0.288018-i 0.0960147 & 0.910951 -i 0.100508  & 
0.309409 -i 0.105268 \\ 
0.04 & 0.827113 -i 0.0885483 & 0.279864-i 0.0925337 & 0.934935 -i 0.103992 & 
0.317891 -i 0.108984 \\ 
0.05 & 0.797636 -i 0.0844276  & 0.269515-i 0.0881547 & 0.966094 -i 0.108558 & 
0.328929 -i 0.113859 \\ 
0.06 & 0.762124 -i 0.0795242 & 0.257076-i 0.0829500 & 1.00467 -i 0.114271 & 
0.342624 -i 0.119965 \\ 
0.07 & 0.720893 -i 0.0739178 & 0.242674-i 0.0770077 & 1.05094 -i 0.121211 & 
0.359095 -i 0.127392 \\ 
0.08 & 0.674327 -i 0.0677029 & 0.226461-i 0.0704316 & 1.10525 -i 0.129471 & 
0.378483 -i 0.136244 \\ 
0.09 & 0.622884 -i 0.0609885 & 0.208617-i 0.0633412 & 1.16797 -i 0.139161 & 
0.400949 -i 0.146644 \\ 
0.1 & 0.567097 -i 0.0538981 & 0.189349-i 0.0558712 & 1.23952 -i 0.150401  & 
0.426674 -i 0.158731 \\ \hline\hline
\end{tabular}%
\caption{Comparison of QNMs of the scalar field by using PT approximation method, and 6th-order WKB approximation method for $\ell=4$, $n=0$, and $m=1$, $Q=0.2$.}
\label{pTellerqnm}
\end{table}

Comparing the PT approximation and the WKB method, we observe that WKB consistently predicts higher real frequencies and stronger damping. Additionally, the presence of a phantom monopole results in larger $\text{Re}(\omega)$ and increased damping compared to the ordinary monopole case, highlighting the influence of $\xi$ on the stability and decay of perturbations.

\section{Grey-body factors}\label{sec6}

\begin{figure}[H]
\begin{minipage}[t]{0.5\textwidth}
        \centering
        \includegraphics[width=\textwidth]{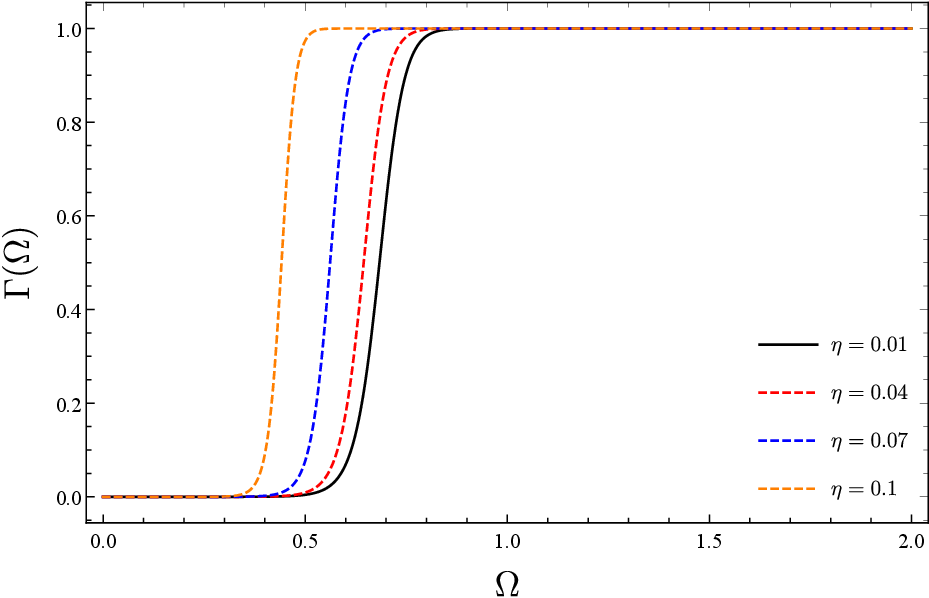}
                \subcaption{$\xi=1$}
        \label{fig:gray1}
\end{minipage}
\begin{minipage}[t]{0.5\textwidth}
        \centering
        \includegraphics[width=\textwidth]{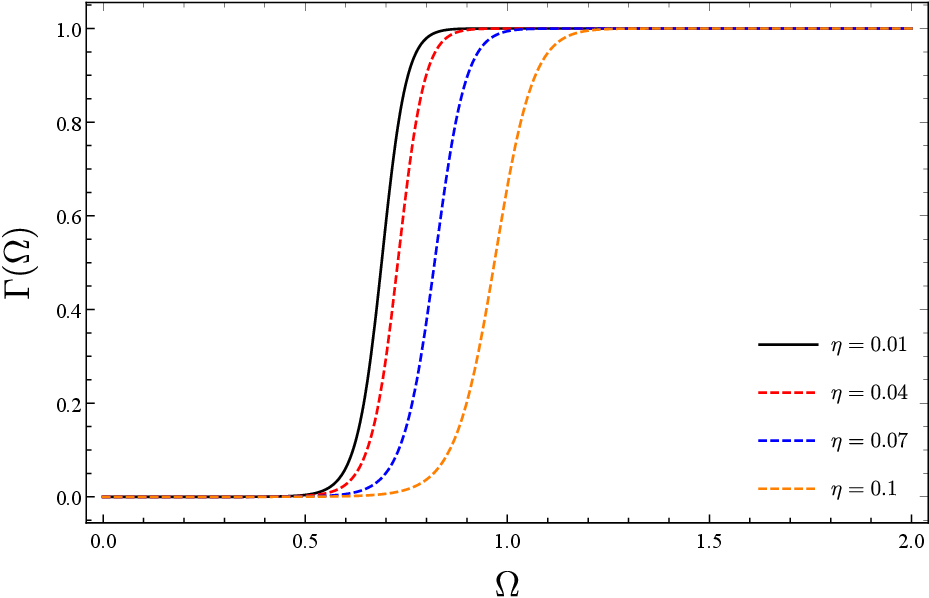}
                \subcaption{$\xi=-1$}
        \label{fig:gray2}
\end{minipage}
\caption{Grey-body factors for $\ell=3$ and $Q=0.1$.}
\label{fig:GBF1}
\end{figure}

In this section, we delve into the grey-body factors which are closely related QNMs \cite{Yang2023, Konoplya:2024lir, Konoplya:2024vuj, Bolokhov:2024otn, Dubinsky:2024vbn, Skvortsova:2024msa}. The grey-body factor represents the probability that radiation emitted close to the event horizon will escape to a distant observer rather than being deflected back by the black hole's gravitational potential \cite{Guo5}. This potential is influenced by the spacetime geometry of the black hole. For the scattering problem, we consider the following wave equation%
\begin{equation}
\frac{d^{2}}{dx^{2}}R(x)+\big( \Omega -\mathcal{V}_{\mathrm{eff}}(x)\big)R(x)=0.
\end{equation}
The boundary conditions used to solve this wave equation are determined by the behavior of the field both close to the event horizon and at infinity.
\begin{itemize}
\item Near the event horizon , $x\rightarrow -\infty $, the field is
entirely ingoing%
\begin{equation}
R(x)\simeq \mathbf{T}e^{-i\Omega x},
\end{equation}%
where $\mathbf{T}$ is the transmission coefficient.

\item At infinity, $x\rightarrow +\infty $, the field consists of a
combination of both ingoing and outgoing waves%
\begin{equation}
R(x)\simeq e^{-i\Omega x}+\mathbf{R}e^{i\Omega x},
\end{equation}%
where $\mathbf{R}$ is the reflection coefficient.
\end{itemize}
These boundary conditions are employed to calculate the intensity of Hawking radiation \cite{NPage1,NPage2,DKokkotas}, or for studying various radiation-related phenomena near black holes, such as superradiance \cite{Zhidenko,RAKonoplya}. In the context of black hole radiation, the transmission coefficient, $\mathbf{T}$, is also known as the grey-body factor. This factor depends on the black hole's spacetime geometry and the properties of the radiation,
\begin{equation}
\Gamma _{\ell }\left( \Omega \right) =\left\vert \mathbf{T}\right\vert
^{2}=1-\left\vert \mathbf{R}\right\vert ^{2}.
\end{equation}%
Grey-body factors and QNMs can be computed using the
higher-order WKB. The formulas were established for spherically symmetric and asymptotically flat black holes by applying the well-known WKB method to calculate grey-body factors,%
\begin{equation}
\Gamma _{\ell }\left( \Omega \right) =\frac{1}{1+e^{i2\pi \mathcal{K}}}.
\end{equation}%
Here $\mathcal{K}$ depends on the values of the effective potential and its derivatives up to the 2i-th order for the third-order WKB method \cite{Schutz:1985km, Iyer:1986np}.  As noted in \cite{Konoplya:2024lir}, the function $\mathcal{K}$ can also be determined through an alternative approach, leveraging its connection to the fundamental mode and the first overtone of the quasinormal spectrum. Therefore, we have:
\begin{eqnarray}
i\mathcal{K} &=&\frac{\Omega ^{2}-\mathrm{Re}\left( \omega _{0}\right) ^{2}}{%
4\mathrm{Re}\left( \omega _{0}\right) \mathrm{Im}\left( \omega _{0}\right) }%
\left( 1+\frac{\big( \mathrm{Re}\left( \omega _{0}\right) -\mathrm{Re}%
\left( \omega _{1}\right) \big) ^{2}}{32\mathrm{Im}\left( \omega
_{0}\right) ^{2}}-\frac{3\mathrm{Im}\left( \omega _{0}\right) -\mathrm{Im}%
\left( \omega _{1}\right) }{24\mathrm{Im}\left( \omega _{0}\right) }\right) 
\notag \\
&-&\frac{\Big( \mathrm{Re}\left( \omega _{0}\right) -\mathrm{Re}\left(
\omega _{1}\right) \Big) }{16\mathrm{Im}\left( \omega _{0}\right) }-\frac{%
\left( \Omega ^{2}-\mathrm{Re}\left( \omega _{0}\right) ^{2}\right) ^{2}}{16%
\mathrm{Re}\left( \omega _{0}\right) ^{3}\mathrm{Im}\left( \omega
_{0}\right) }\Bigg(1-\frac{\mathrm{Re}\left( \omega _{0}\right) \big( \mathrm{Re}\left(
\omega _{0}\right) -\mathrm{Re}\left( \omega _{1}\right) \big) }{4\mathrm{%
Im}\left( \omega _{0}\right) ^{2}}\Bigg)  \notag \\
&+&\frac{\left( \Omega ^{2}-\mathrm{Re}\left( \omega _{0}\right) ^{2}\right)
^{3}}{32\mathrm{Re}\left( \omega _{0}\right) ^{5}\mathrm{Im}\left( \omega
_{0}\right) }\Bigg[1+\frac{\mathrm{Re}\left( \omega _{0}\right) \left( 
\mathrm{Re}\left( \omega _{0}\right) -\mathrm{Re}\left( \omega _{1}\right)
\right) }{4\mathrm{Im}\left( \omega _{0}\right) ^{2}}  \notag \\
&+&\mathrm{Re}\left( \omega _{0}\right) ^{2}\left( \frac{\left( \mathrm{Re}%
\left( \omega _{0}\right) -\mathrm{Re}\left( \omega _{1}\right) \right) ^{2}%
}{16\mathrm{Im}\left( \omega _{0}\right) ^{4}}-\frac{3\mathrm{Im}\left(
\omega _{0}\right) -\mathrm{Im}\left( \omega _{1}\right) }{12\mathrm{Im}%
\left( \omega _{0}\right) }\right) \Bigg]+\mathcal{O}\left( \frac{1}{\ell ^{3}%
}\right) .
\end{eqnarray}
Here, $\omega_{0}$ is the the fundamental mode and $\omega_{1}$ is the first overton.
Figure~\ref{fig:GBF1} illustrates the grey-body factors as a function of frequency $\Omega$ for a nonlinear magnetic-charged black hole with both ordinary ($\xi = 1$) and phantom ($\xi = -1$) global monopoles. 

\noindent The grey-body factor represents the probability that radiation emitted near the event horizon escapes to infinity rather than being reflected back by the potential barrier. As seen in the plot, the grey-body factor increases with frequency, meaning that higher-energy modes have a higher probability of transmission. Additionally, the presence of a phantom global monopole ($\xi = -1$) leads to a lower transmission coefficient compared to the ordinary case, indicating that the phantom field modifies the effective potential in a way that enhances reflection and suppresses radiation escape. 

\begin{figure}[H]
\begin{minipage}[t]{0.5\textwidth}
        \centering
        \includegraphics[width=\textwidth]{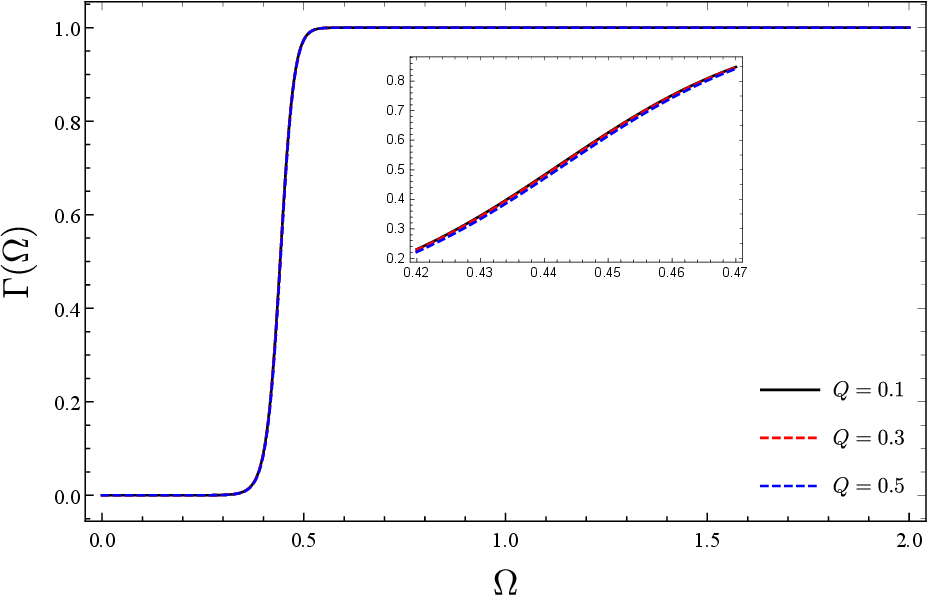}
                \subcaption{$\xi=1$}
        \label{fig:gray11}
\end{minipage}
\begin{minipage}[t]{0.5\textwidth}
        \centering
        \includegraphics[width=\textwidth]{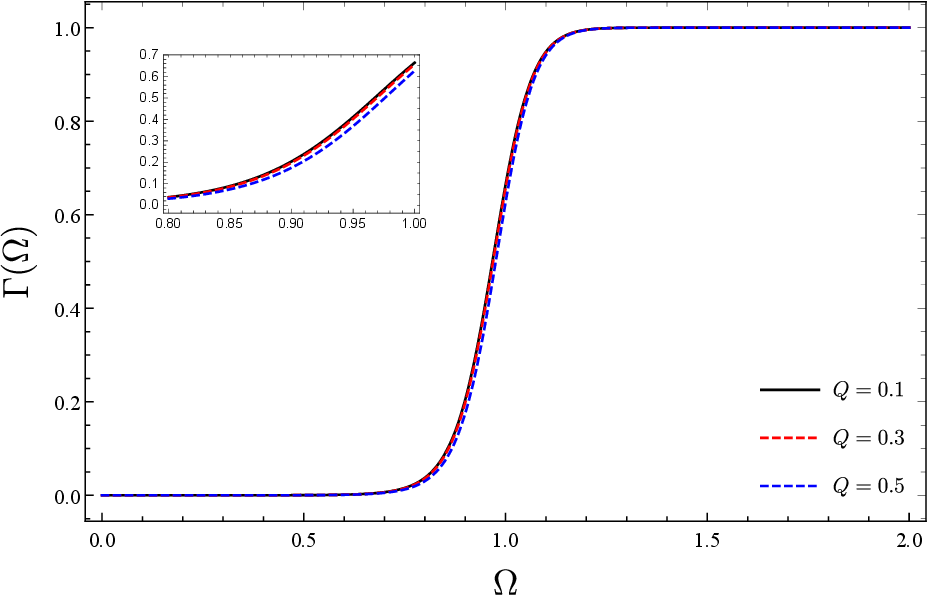}
                \subcaption{$\xi=-1$}
        \label{fig:gray22}
\end{minipage}
\caption{Grey-body factors for $\ell=3$ and $\eta=0.1$.}
\label{fig:GBF2}
\end{figure}

In Figure~\ref{fig:GBF2}, we examine the dependence of the grey-body factors on the symmetry-breaking parameter $\eta$ for black holes with both ordinary and phantom global monopoles. 

\noindent The results indicate that as $\eta$ increases, the transmission probability decreases, leading to a suppression of radiation escaping to infinity. This effect is more pronounced in the phantom case ($\xi = -1$), suggesting that stronger symmetry breaking increases the effective potential barrier surrounding the black hole. This result aligns with our findings in thermodynamics, where increasing $\eta$ leads to modifications in black hole stability and temperature, further influencing the radiation emission process.

\begin{figure}[H]
\begin{minipage}[t]{0.5\textwidth}
        \centering
        \includegraphics[width=\textwidth]{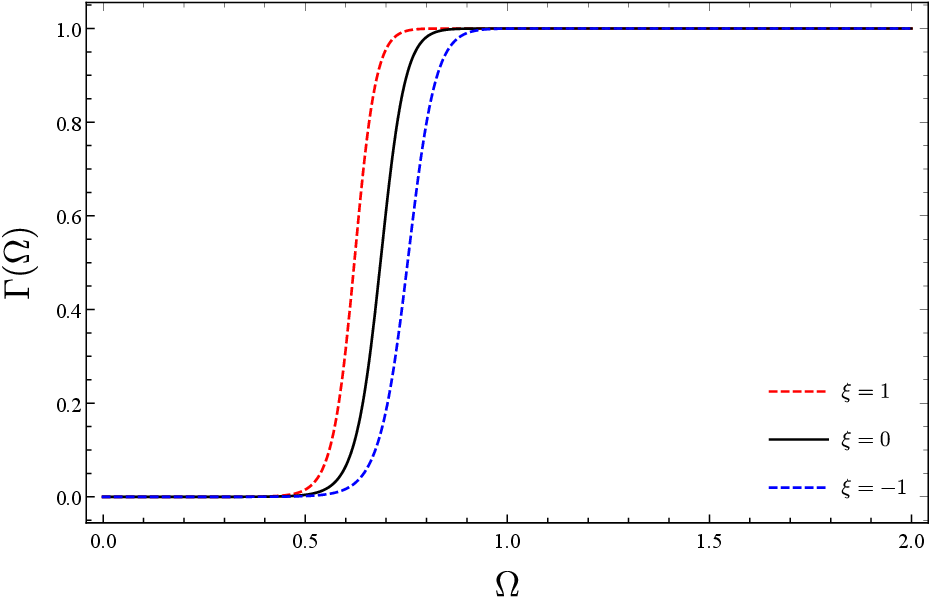}
                \subcaption{$\eta=0.05$}
        \label{fig:gray111}
\end{minipage}
\begin{minipage}[t]{0.5\textwidth}
        \centering
        \includegraphics[width=\textwidth]{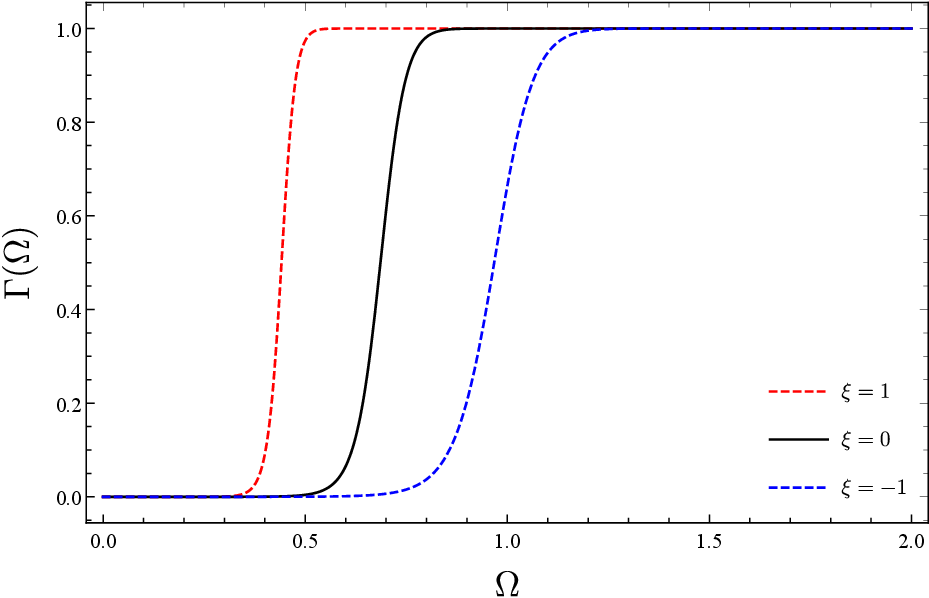}
                \subcaption{$\eta=0.11$}
        \label{fig:gray222}
\end{minipage}
\caption{Grey-body factors for $\ell=3$, $\eta=0.1$ and $Q=0.2$.}
\label{fig:GBF3}
\end{figure}

Figure~\ref{fig:GBF3} explores the effect of the nonlinear magnetic charge $Q$ on the grey-body factors.

\noindent As seen in the figure, increasing $Q$ leads to a reduction in the transmission coefficient, meaning that a stronger magnetic charge suppresses the escape of radiation. This behavior is expected since a larger $Q$ enhances the curvature effects and modifies the effective potential, making it more difficult for radiation to overcome the potential barrier. Notably, this effect is consistent across both ordinary and phantom monopole cases, reinforcing the role of NLE in shaping black hole radiation characteristics.

\section{Conclusion}\label{sec7}

In this work, we investigated the properties of a nonlinear magnetic-charged black hole in the presence of a phantom global monopole. By incorporating nonlinear electrodynamics (NLE) and exotic scalar fields, we derived an exact black hole solution and analyzed its thermodynamic and dynamical properties. Our findings reveal significant deviations from standard general relativity (GR), providing potential observational signatures for modified gravity theories. 

We examined the black hole’s thermodynamic properties, including its Hawking temperature, entropy, and heat capacity. The presence of the phantom global monopole was shown to modify the thermodynamic stability, leading to novel phase structures. Notably, the Hawking temperature exhibits a dependence on the monopole parameter, influencing the evaporation process of the black hole. The heat capacity analysis indicated the presence of a second-order phase transition, suggesting regions of stability and instability in the black hole's thermodynamic phase space. Although the influence of the phantom monopole may appear as an overall shift in the temperature and heat capacity curves, this shift encodes significant physical effects: the conical excess introduced by the phantom field alters the asymptotic geometry, reduces the black hole temperature at fixed horizon radius, and shifts the location of phase transition points. These changes reflect a genuine modification of the black hole’s thermodynamic behavior due to the topological deformation introduced by the scalar field.

We also analyzed the geodesic structure of test particles, exploring their motion in the modified black hole background. The analysis reveals that both the phantom global monopole and the nonlinear magnetic charge contribute significantly to the structure and stability of circular orbits. Notably, the presence of the monopole tends to shift the innermost stable circular orbit (ISCO) outward and enhance orbital instability, while the NLE acts to stabilize orbits at smaller radii. The Lyapunov exponent, which measures the instability timescale of orbits, increases with the symmetry-breaking parameter $\eta$, mirroring the faster damping rates observed in the quasinormal mode spectrum. These findings not only reinforce the consistency between the dynamical and perturbative analyses, but also suggest possible observational signatures in accretion disk dynamics and gravitational wave signals near such exotic black holes. 
Furthermore, we studied the quasinormal modes (QNMs) of the black hole, employing both the sixth-order WKB approximation and the Pöschl-Teller potential method. The QNM spectrum provided insights into the stability of the system under perturbations. Our results showed that the presence of NLE and the phantom global monopole modifies the real and imaginary parts of the QNM frequencies, affecting the damping times of perturbations. 

A key addition to our study was the analysis of grey-body factors, which describe the probability of radiation escaping the black hole’s gravitational potential. We found that the presence of the phantom global monopole reduces the transmission coefficient, leading to a suppression of escaping radiation. This effect was further influenced by the symmetry-breaking parameter and the magnetic charge, reinforcing the role of NLE in shaping black hole radiation properties.

Overall, our findings highlight the intricate interplay between NLE, phantom fields, and global monopoles in black hole physics. The modifications introduced by these exotic fields have direct implications for astrophysical observations, including the shadows of black holes, gravitational waves, and Hawking radiation. Future work could extend this analysis by exploring rotating solutions, higher-dimensional generalizations, or the impact of additional modified gravity corrections.

\section*{Conflict of Interests} 

The authors declare no such conflict of interest.

\section*{Data Availability Statement} 

No data were generated or analyzed in this study.

\section*{Acknowledgments}
{The authors thank the anonymous reviewers for their constructive comments and valuable suggestions, which greatly improved the quality of this work.} B. C. L. is grateful to Excellence Project FoS UHK 2205/2025-2026 for the financial support.

\end{document}